\documentclass[smallcondensed]{svjour3}     % onecolumn (ditto)
\usepackage{graphicx}
\usepackage{geometry}
\usepackage{amsmath}
\usepackage{amsfonts}
\usepackage{natbib}
\usepackage{hyperref}
\hypersetup{colorlinks=true, citecolor=blue, linkcolor=blue}

\def \vv#1{\mathbf{#1}}
\def \be {\begin{equation}}
\def \ee {\end{equation}}

\def\crm{\cr\noalign{\medskip}}

\def\ii {\mathrm{i}}
\def\Fp {F_{+}}
\def\Fm {F_{-}}
\def\Gc {{\cal G}}
\def\Ms {m_0}
\def\ms {m}
\def \om {\omega}
\def \up {\upsilon}
\def \vpi {\varpi}
\def \ek {\phi}
\def \cT {x}
\def \sT {\sin \theta}
\def \es {y}
\def \eks {z}

\def\xii {\hat x_1}
\def\xj {\hat x_2}
\def\xk {\hat x_3}

\def \ve {\vv{e}}
\def \ue {\hat \ve}
\def \de {\dot \ve}
\def \vr {\vv{r}}
\def \ur {\hat \vr}

\def \vk {\vv{k}}
\def \vx {\vk \times \ue}
\def \vp {\vv{p}}
\def \vq {\vv{q}}
\def \vw {\boldsymbol{\om}}
\def \vs {\vv{s}}
\def \vu {\vv{u}}
\def \vL {\vv{L}}
\def \vF {\vv{F}}
\def \vG {\vv{G}}
\def \vR {\vv{R}}
\def \vT {\vv{T}}
\def\TI {{\bf \cal I}}
\def\SR {{\bf \cal S}}

\def \kl {{k'}}
\def \sumk {\sum_{k=-\infty}^{+\infty}}
\def \sumkl {\sum_{\kl=-\infty}^{+\infty}}
\def \kf {k_{\rm f}}
\def \tu {\tau}
\def \te {\tau_e}
\def \ta {\tau_a}
\def \tv {\tau_v}
\def \sq {\sqrt{1-e^2}}
\def \Ae {{\cal E}_0}
\def \At {{\cal T}_0}
\def \Ke {{\cal K}_e}
\def \Kt {{\cal K}_t}
\def \az {a(-kn)}
\def \au {a(\om-kn)}
\def \ad {a(2\om-kn)}
\def \bz {b(-kn)}
\def \bu {b(\om-kn)}
\def \bd {b(2\om-kn)}
\def \Xtz {X_k^{-3,0}}
\def \Xtup {X_k^{-3,1}}
\def \Xtum {X_k^{-3,-1}}
\def \Xtdp {X_k^{-3,2}}
\def \Xtdm {X_k^{-3,-2}}
\def\ep {\mathrm{e}}
\def \eq {e}
\def \pt {p}
\def\tl {t'}

\def\iinf {\int}
\def\spi{}
\def\sqpi{}

\def\bibpath{}
\def \llabel#1{\label{#1}}
\def\bfx#1{#1}
\begin{document}
\title{Tidal evolution for any rheological model using a vectorial approach expressed in Hansen coefficients.}

\titlerunning{Tidal evolution using a vectorial approach}        % if too long for running head

\author{Alexandre C. M. Correia         \and    Ema F. S. Valente}

\authorrunning{A.C.M. Correia \and E.F.S. Valente} % if too long for running head

\institute{A. C. M. Correia 
	%\email{alexandre.correia@uc.pt}
        \at CFisUC, Departamento de F\'isica, Universidade de Coimbra, 3004-516 Coimbra, Portugal
        \at IMCCE, Observatoire de Paris, PSL Universit\'e, 77 Av. Denfert-Rochereau, 75014 Paris, France
	\and
	E. F. S. Valente
        \at CFisUC, Departamento de F\'isica, Universidade de Coimbra, 3004-516 Coimbra, Portugal\\
}

\date{\today}

\maketitle

\begin{abstract}
We revisit the two body problem, where one body can be deformed under the action of tides raised by the companion.
Tidal deformation and consequent dissipation result in spin and orbital evolution of the system. 
In general, the equations of motion are derived from the tidal potential developed in Fourier series expressed in terms of Keplerian elliptical elements, so that the variation of dissipation with amplitude and frequency can be examined.
However, this method introduces multiple index summations and some orbital elements depend on the chosen frame, which is prone to confusion and errors. % in the equations of motion.
Here, we develop the quadrupole tidal potential solely in a series of Hansen coefficients, which are widely used in celestial mechanics and depend just on the eccentricity.
We derive the secular equations of motion in a vectorial formalism, which is frame independent and valid for any rheological model.
We provide expressions for a single average over the mean anomaly and for an additional average over the argument of the pericentre.
These equations are suitable to model the long-term evolution of a large variety of systems and configurations, from planet-satellite to stellar binaries.
We also compute the tidal energy released inside the body for an arbitrary configuration of the system.


\end{abstract}

%________________________________________________________________

%\tableofcontents

\section{Introduction}

Tidal effects arise when an extended body is placed in a differential gravitational field, such as the one generated by a point mass companion.
For non-rigid bodies, each mass element adjusts to the equipotential surface, which gives rise to a global redistribution of mass known as tidal bulge.
In the process, the friction inside the extended body introduces a delay between the initial perturbation and the maximal deformation. 
As a consequence, the companion exerts a torque on the tidal bulge of the extended body, which modifies its spin and the orbit.

The estimates for the %spin and orbital
tidal evolution of a body are based on a very general formulation of the tidal potential, initiated by \citet{Darwin_1879a}.
Assuming a homogeneous body consisting of an incompressible fluid with constant viscosity,  \citeauthor{Darwin_1880} derived a tide-generated disturbing potential expanded into a Fourier series expressed in terms of elliptical elements.
The equations of motion are then obtained using the Lagrange planetary equations \citep{Darwin_1880}.

\citet{Kaula_1964} writes the tidal potential using Love numbers \citep{Love_1911}. 
Each term of the Fourier series involves a Love number associated with the amplitude of the tide and a phase lag accounting for the non-instantaneous deformation of the body.
This description is more general than \citeauthor{Darwin_1880}'s, because it makes no assumption on the rheology of the body \citep[e.g.,][]{Efroimsky_2012b}.
Different rheologies have been proposed for satellites, rocky planets, giant gaseous or stars \citep[e.g.,][]{Mignard_1979, Ogilvie_Lin_2004,Ogilvie_Lin_2007, Efroimsky_Lainey_2007, Ferraz-Mello_2013, Correia_etal_2014, Renaud_Henning_2018},
some because of their simplicity, others motivated by theoretical studies, laboratory experiments or geophysical measurements.

The description by \citeauthor{Kaula_1964} also has a more compact and systematic form of the tidal potential, thus being more convenient for examining the tidal effects of varying conditions. 
However, sometimes there is an ambiguity in the interpretation of the frequencies involved in the phase lags \citep[e.g.,][]{Efroimsky_Williams_2009}.
Moreover, mistakes such as neglecting energy or momentum conservation considerations are more easily made.
Indeed, \citet{Boue_Efroimsky_2019} have shown that additional terms need to be included owing to the conservation of the angular momentum and non-inertial frame considerations.

The classical expansion of the tidal potential in elliptical elements depends on the chosen frame and also introduces multiple index summations, which can lead to confusion and errors in the equations of motion.
In the case of a linear constant time-lag model, which can be seen as a first order expansion of a viscoelastic rheology \citep{Singer_1968, Mignard_1979}, it has been shown that the equations of motion are more easily expressed in terms of angular momentum vectors \citep[e.g.,][]{Correia_2009, Correia_etal_2011}.
Therefore, in this paper we aim to also use these vectors for any rheological model that can be expressed in terms of Love numbers.
This formalism is independent of the reference frame and allows to simply add the tidal contributions of multiple bodies in the system.

In Sect.\,\ref{tbpwt}, using the quadrupole tidal potential expanded in series of Hansen coefficients, we obtain the equations of motion using vectors and Love numbers.
In Sect.\,\ref{singav} and Sect.\,\ref{doubav}, 
we average the equations of motion over the mean anomaly, and over the argument of the pericentre, respectively,
which provide simpler expressions that are easily to implement and suitable for long-term evolution studies.
In Sect.\,\ref{planarcase}, we provide the equations of motion for the more simple planar case, and in Sect.\,\ref{linearapprox}, we explain how the equations of motion simplify for the linear constant time-lag model.
Finally, we discuss our results in Sect.\,\ref{sectconc}.

\section{Two body problem with tides}

\llabel{tbpwt}

We consider a system of two bodies with masses $\Ms$ and $\ms$ in a Keplerian orbit.
The orbital angular momentum is given by \citep[e.g.,][]{Murray_Dermott_1999}
\be 
\vG = \beta \sqrt{\mu a (1-e^2)} \, \vk 
\ , \label{210804a}
\ee
where $a$ is the semi-major axis, $e$ is the eccentricity,
$\beta = \Ms \ms / (\Ms + \ms)$,
$\mu = \Gc (\Ms + \ms)$,
$\Gc$ is the gravitational constant,
and $\vk$ is the unit vector along the direction of $\vG$, which is normal to the orbit.
The body with mass $\Ms$, named as ``perturber'', is a point mass object.
The body with mass $\ms$, named as ``central body'', \bfx{is an extended body} and can be deformed under the action of tides.
It rotates with angular velocity $\vw = \om \, \vs$, where $\vs$ is the unit vector along the direction of the spin axis.
The rotational angular momentum is given by
\be
\vL = C \vw + \TI \cdot \vw  
\ , \llabel{210804b}
\ee
where $C$ is the moment of inertia of a sphere and 
\be
\TI = 
\left[\begin{array}{rrr} 
I_{11}&  I_{12}& I_{13} \\
I_{12}&  I_{22}& I_{23} \\
I_{13}&  I_{23}& I_{33} 
\end{array}\right]
\label{121026c}
\ee
is the inertia tensor that accounts for the departure from the sphere.
In the absence of the perturber, the central body is a perfect sphere, and so $\TI=0$.
However, when the perturber is present, the central body is deformed, and $\TI$ can change.
\bfx{In this work we assume that the central body is incompressible, and thus $\mathrm{tr}(\TI)=I_{11} + I_{22} + I_{33} =0$ \citep[e.g.,][]{Rochester_Smylie_1974}.}
%aqui

In general, 
%for planets and stars the rotation about the axis of maximal inertia is usually much faster than any change in this axis’ orientation. We can then average over the rotation angle (gyroscopic approximation), \bfx{and the rotation axis becomes the same as the axis of figure} \citep[e.g.,][]{Boue_Laskar_2006}. Moreover, 
tidal deformations are small and yield periodic changes in the moments of inertia, such that $\dot \TI \vw \ll C \dot \vw$ \citep[e.g.,][]{Frouard_Efroimsky_2018}. % Sect. 3.1
Therefore, for simplicity, we can assume that the deformation is small with respect to the radius of the unperturbed sphere, $R$, such that $I_{ij} \ll C$ ($i,j=1,2,3$), and
\be
\vL \approx C  \vw  = C \om \, \vs
% \quad \Leftrightarrow \quad \vw \approx \frac{\vL}{C} 
\ .  \llabel{151019b}
\ee

\subsection{Potential energy of a non-spherical body}

The gravitational potential of the central body at a generic position $\vr$ from its centre of mass is given by \citep[e.g.,][]{Goldstein_1950}
\be 
V (\vr) = - \frac{\Gc \ms}{r} + \Delta V (\vr)
\, \llabel{210805a}
\ee
with
\be
%\Delta V (\vr) = \frac{3 \Gc}{2 r^3} \left[ \ur \cdot \TI \cdot \ur - \frac{1}{3} \mathrm{tr}(\TI) \right] \ , \llabel{121026b}
\Delta V (\vr) = \frac{3 \Gc}{2 r^3} \, \ur \cdot \TI \cdot \ur  \ , \llabel{121026b}
\ee
where $r = ||\vr||$ is the norm, $\ur = \vr / r =  (\xii, \xj, \xk)$ is the unit vector, %$\mathrm{tr}(\TI) = I_{11} + I_{22} + I_{33} $, %aqui
and we neglect terms in $(R/r)^3$ (quadrupolar approximation).

The point mass $\Ms$ interacts with the potential of the central body (Eq.\,(\ref{210805a})).
The non-spherical contribution of this potential, $\Delta V (\vr)$, is responsible for the perturbations to the Keplerian motion.
The corresponding potential energy, $ U (\vr) = \Ms \Delta V (\vr) $, can be rewritten as
\be
\begin{split}
U (\vr) =
\frac{3 \Gc \Ms}{2r^3}  \Big[  \big(I_{22}-I_{11}\big) \big(\xj^2 - \frac13\big) + \big(I_{33}-I_{11}\big) \big(\xk^2 -\frac13\big) %\\
 + \, 2 \big( I_{12} \xii \xj + I_{13} \xii \xk + I_{23} \xj \xk \big) \Big]
\ . \llabel{191014a}
\end{split}
\ee

\subsection{Reference frames}

In this work we use two different reference frames, one attached to the orbit of the perturber and another to the spin axis of the central body (see Fig.~\ref{frames}).
The reason for this choice is that the unit vectors $\vk$ and $\vs$ are easily directly obtained from the orbital (Eq.\,(\ref{210804a})) and rotational (Eq.\,(\ref{151019b})) angular momentum vectors, respectively.
In general, these two vectors are not \bfx{collinear}, and so in a first step we need to build two independent frames as in \citet{Correia_2006}.

\begin{figure}
\begin{center}
\includegraphics[width=9.2cm]{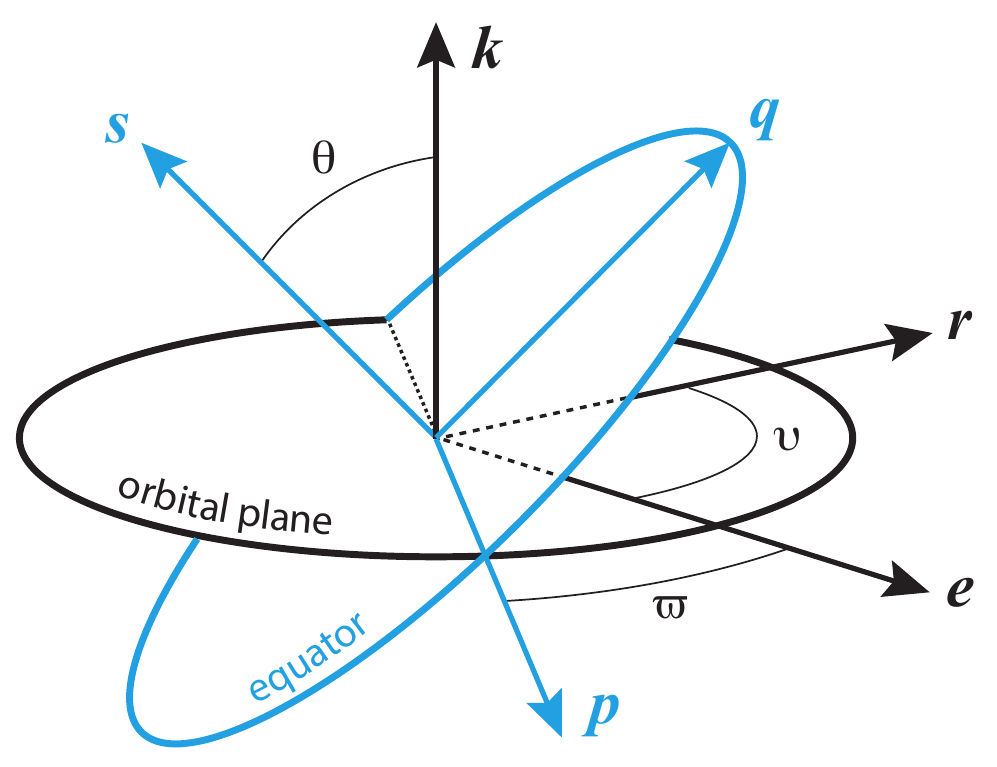} 
 \caption{Reference frames and angles. $(\ue, \vx, \vk)$ is a cartesian frame such that $\vk$ is a unit vector normal to the orbital plane, $\ve$ is the Laplace-Runge-Lenz vector (where $\ue = \ve / e$ gives the direction of the pericentre), $\vpi$ is the argument of the pericentre and $\up$ is the true anomaly. ($\vp,\vq,\vs$) is a cartesian frame, such that $\vs$ is a unit vector normal to the equatorial plane of the central body, $\vp$ is a unit vector along the line of nodes of the two reference planes, and $\theta$ is the angle between them. Note that, although $\vs$ gives the direction of the spin axis, the vectors $\vp$ and $\vq = \vs \times \vp$ follow the orbital plane and not the rotation. \llabel{frames}  }
\end{center}
\end{figure}

We let $(\ue, \vx, \vk)$ to be a cartesian frame such that $\vk$ is aligned with the normal to the orbit, and $\ue = \ve / e$ is aligned with the pericentre of the orbit and obtained from the Laplace vector
\be
\ve = \frac{\dot \vr \times \vG}{\beta \mu} - \ur 
\ . \llabel{210805d}
\ee

We further let ($\vp,\vq,\vs$) to be a cartesian frame, such that
\be
\vp = \frac{ \vk \times \vs }{\sin \theta} \ , %\quad \mathrm{and} 
\quad \vq = \vs \times \vp = \frac{\vk - \cos \theta \, \vs}{\sin \theta}  \ , 
%\quad \vs = \vp \times \vq \ ,
\ee
where $\vs$ is aligned with the spin axis, and $\vp$ is aligned with the line of nodes between the equator of the central body and the orbital plane of the perturber (Fig.~\ref{frames}).
The angle $\theta$ corresponds to the angle between $\vk$ and $\vs$, that is, $\cos \theta = \vk \cdot \vs$.

These two frames are also known as ``precession frames'',  as they follow the precession of the pericentre and node, respectively.
They are connected through the rotation matrix
\be
\left[\begin{array}{c}  
\ue  \\ \vx  \\ \vk
\end{array}\right] 
%= {\cal R}_3 (-\vpi) {\cal R}_1 (\theta)
= \left[\begin{array}{ccccc}  
\cos \vpi && \cos \theta \sin \vpi && - \sin \theta \sin \vpi  \\ 
- \sin \vpi && \cos \theta \cos \vpi && - \sin \theta \cos \vpi  \\ 
0 && \sin \theta && \cos \theta  
\end{array}\right] 
\left[\begin{array}{c}  
\vp  \\ \vq  \\ \vs
\end{array}\right] 
\ , \llabel{210804d}
\ee
where $\vpi$ is the argument of the pericentre.
We note that these two frames are not inertial, because the orbit and the spin can evolve due to tidal interactions.
However, \bfx{the changes are very small during an orbital %and rotational 
period, and so the two frames can be assumed as constant}.

%$$ \ue = (\ue \cdot \vp) \, \vp + (\ue \cdot \vq) \, \vq + (\ue \cdot \vs) \, \vs = \cos \vpi \, \vp + \cos \theta \sin \vpi \, \vq - \sin \theta \sin \vpi \, \vs $$

The position vector can be expressed in these two frames as
\be
\ur =  \cos \up \, \ue + \sin \up \, \vx = \xii \, \vp + \xj \, \vq + \xk \, \vs 
\ , \label{210804c}
\ee
where $\up$ is the true anomaly, and
\be
\begin{split}
\xii & = \ur \cdot \vp = \cos \up \, \ue \cdot \vp + \sin \up \, \frac{\ue \cdot \vs }{\sin \theta} = \cos (\vpi+\up) \ , \crm
\xj & = \ur \cdot \vq = - \cos \up \, \frac{\cos \theta \, \ue \cdot \vs}{\sin \theta} + \sin \up \cos \theta \, \ue \cdot \vp =  \cos \theta \sin (\vpi+\up) \ , \crm
\xk & = \ur \cdot \vs = \cos \up \, \ue \cdot \vs - \sin \up \sin \theta \, \ue \cdot \vp
= - \sin \theta \sin (\vpi+\up) \ .
\end{split}
 \label{210804e}
\ee
Similarly,
\be
\begin{split}
\dot \vr = \frac{n a}{\sqrt{1-e^2}} \, \vk \times \left( \ur + \ve \right) & =
 \frac{n a}{\sqrt{1-e^2}} \,  \big( (\cos \up + e) \, \vk \times \ue - \sin \up \, \ue ) \big)  \ ,
% \\ & = \dot \ix \, \vp + \dot \iy \, \vq + \dot \iz \, \vs 
\end{split}
\ee
with
\be
\begin{split}
\dot \vr \cdot \vp & =
\frac{- n a}{\sqrt{1-e^2}} \,  \big( \sin ( \vpi +  \up ) + e \sin \vpi \big) \ , \crm
\dot \vr \cdot \vq & =
 \frac{n a}{\sqrt{1-e^2}} \cos \theta \, \big( \cos (\vpi + \up) + e \cos \vpi \big) \ , \crm
\dot \vr \cdot \vs & =
 \frac{-n a}{\sqrt{1-e^2}}  \sin \theta \, \big( \cos (\vpi + \up) + e \cos \vpi \big) \ ,
\end{split}
\ee
where $n = \sqrt{\mu / a^3}$ is the mean motion.

\subsection{Equations of motion}
\llabel{pmp}

The tidal force between the two bodies is easily obtained from the potential energy (Eq.\,(\ref{191014a})) as
\be
\vF = - \nabla U (\vr) =  \vF_1 + \vF_2
\ , \llabel{170911d}
\ee
with

\be
%\begin{split}
\vF_1  = %&
\frac{15 \Gc \Ms}{r^4}  \Big[ \frac{I_{22}-I_{11}}{2} \big(\xj^2 - \frac15 \big)  
+ \frac{I_{33}-I_{11}}{2} \big(\xk^2 - \frac15 \big) %\\ &
 + I_{12} \xii \xj + I_{13} \xii \xk + I_{23} \xj \xk \Big] \ur 
\ , \llabel{170911b}
%\end{split}
\ee
and
\be
\begin{split}
\vF_2 = &
 -  \frac{3 \Gc \Ms}{r^4}  \Big[ \big(I_{22}-I_{11}\big) \xj \, \vq + \big(I_{33}-I_{11}\big) \xk \, \vs  \\ 
 & + I_{12} (\xii \, \vq + \xj \, \vp)  + I_{13} (\xii \, \vs + \, \xk \, \vp) + I_{23} (\xj \, \vs + \xk \, \vq) \Big]
\ . \llabel{170911c}
\end{split}
\ee
We decompose $\vF$ and the following vectorial quantities in the frame $(\vp, \vq, \vs)$, as this greatly facilitates the computation of the inertia tensor $\TI$ (Sect.~\ref{deformation_body}).
%Therefore, all the coefficients $I_{ij}$ (with $i,j = 1,2,3$) correspond to a projection in this basis.
There is no loss of generality with this choice, because the vectors $(\vp, \vq, \vs)$ can always be expressed in another basis.
In particular, we follow the evolution of the system in an inertial frame, since nothing forbids to project an inertial vector on a non-inertial coordinate system.
We obtain for the orbital evolution of the system 
\be
\ddot \vr =  - \frac{\mu}{r^2} \, \ur + \frac{\vv{F}}{\beta} \ .  \llabel{151028c}
\ee
The first term corresponds to the Keplerian motion, while the second one corresponds to the correction from the tidal force. 
From a secular evolution perspective, we only care about the modifications in the orbit and spin that are brought by the second term.
Therefore, the evolution of the systems is better described by the evolution of $\vG$ and $\ve$ for the orbit and $\vL$ for the spin.

The evolution of the angular momentum vectors is computed from the gravitational torque. In an inertial frame, we have: 
\be
\dot \vG = \vT = \vr \times \vv{F} = \vr \times \vF_2 
\ , \llabel{150626a}
\ee
and, owing to the conservation of the total angular momentum,
\be
\dot \vL = - \dot \vG = - \vT
\ , \llabel{210805b}
\ee
with
\be
\llabel{151028e}
\begin{split}
\vT = - \frac{3 \Gc \Ms}{r^3} \bigg\{ &
\left[ \big(I_{33}-I_{22}\big) \xj \xk - I_{12} \xii \xk + I_{13} \xii \xj  + I_{23} (\xj^2 - \xk^2) \right] \vp \\
+ &  \left[ \big(I_{11}-I_{33}\big) \xii \xk + I_{12} \xj \xk + I_{13} (\xk^2 - \xii^2) - I_{23} \xii \xj \right] \vq \\
+ & \left[ \big(I_{22}-I_{11}\big) \xii \xj + I_{12} (\xii^2 - \xj^2) - I_{13} \xj \xk + I_{23} \xii \xk \right] \vs \, \bigg\} \ . 
\end{split}
\ee
%%%%%%%%%%%%%%%%

The evolution of the Laplace vector is obtained by differentiating expression (\ref{210805d}) 
\be
\dot \ve = \frac{1}{\beta \mu} \left( \frac{\vF}{\beta} \times \vG + \dot \vr \times \vT  \right) 
\ . \llabel{210805c}
\ee

\subsection{Deformation of the central body}

The mass distribution inside the central body is characterised by the coefficients of the inertia tensor, $\TI$ (Eq.\,(\ref{121026c})).
It is the result of self gravity and the response to any perturbing potential.
Here, we consider that the central body is \bfx{an extended body} that deforms solely under the action of the differential gravitational force from the companion body of mass $\Ms$. 
A mass element $d m$ at a location $\vr'$  is thus subject to the perturbing potential \citep[e.g.][]{Lambeck_1980}
\be
V_\pt (\vr') = -\frac{\Gc \Ms}{r} \left(\frac{r'}{r}\right)^2 P_2 (\ur' \cdot \ur) 
\ , \llabel{121026e}
\ee
where  $\vr$ is the position of the mass $\Ms$ with respect to the centre of mass of the central body, $P_2(x) = (3 x^2 -1)/2 $ is a Legendre polynomial, and we neglected terms higher than $(r'/r)^3$, that is, we consider only the quadrupole perturbations.
Thus, on the central body's surface, $\vR$, % the harmonic part
the non-spherical contribution of the perturbing potential is given by 
\be
V_\pt (\vR) = -\frac{\Gc \Ms}{r} \left(\frac{R}{r}\right)^2 P_2 (\hat \vR \cdot \ur) 
\ , \llabel{210908a}
\ee
where $R = |\vR|$.
For simplicity, since the surface of the central body is nearly spherical, we can assume $R$ to be constant and equal to the average radius.

The above potential can be rearranged as
\be
V_\pt (\vR) %=  \frac{3 \Gc}{2 R^3} \left[ \hat \vR \cdot \TI_\pt \cdot \hat \vR -\frac{1}{3} \mathrm{tr}(\TI_\pt) \right]  
= \frac{3 \Gc}{2 R^3} \, \hat \vR \cdot \TI_\pt \cdot \hat \vR 
\ , \llabel{171110x}
\ee
with
\be
\frac{\TI_\pt}{\ms R^2} = 
-  \frac{\Ms}{\ms} \left( \frac{R}{r } \right)^3  \left( \ur  \, \ur^T - \frac{1}{3}  \mathbb{I} \right)
\ , \llabel{121030b}
\ee
where $^T$ denotes the transpose, and $\mathbb{I}$ is the identity matrix.
The ``perturbing'' tensor $\TI_\pt$ can be seen as a perturbation of the inertia tensor, $\TI$  (Eq.\,(\ref{121026c})). %, with $\mathrm{tr}(\TI_\pt)=0$.

A convenient way of handling the deformation is through the Love number approach \citep[e.g.][]{Love_1927, Peltier_1974}. 
It assumes that the deformations are small and can be made linear, that is, proportional to the perturbing potential.
The total tidal potential is thus a linear combination of the tidal responses to each excitation $V_\pt (\vR, \tl)$ over time, with $\tl \le t$.
In the frame of the central body, $B$, for any point of its surface we thus have 
\be
\Delta V (\vR, t) = \sqpi \iinf k_2(t-\tl) V_\pt (\vR, \tl) d \tl = k_2(t) *  V_\pt (\vR, t) 
\ , \llabel{210909a} 
\ee
where $*$ is the convolution product. 
$k_2(t)$ is a {\it Love distribution} such that $k_2(t) = 0$ for all $t<0$, % (by analogy with the Love number $k_2$) 
which depends on the internal structure of the central body \citep[for more details see][]{Boue_etal_2016, Boue_etal_2019p}.
\bfx{The knowledge of $\Delta V (\vR, t)$ is sufficient to constrain $\Delta V (\vr, t)$, because it is the solution of a Dirichlet problem.}
Therefore, by comparing expressions (\ref{171110x}) and (\ref{210909a}) with the gravitational potential (Eq.\,(\ref{121026b})) taken at the central body's surface ($\vr = \vR$), we finally get for the deformation
\be
\TI^B(t) = \sqpi \iinf k_2(t-\tl) \, \TI^B_p(\tl) \, d \tl  = k_2(t) * \TI^B_p(t) 
% = ( k_2 * \TI^B_p) (t) 
\ , \llabel{200116a} 
\ee
where $\TI^B$ and $\TI^B_\pt$ are the inertia (Eq.\,(\ref{121026c})) and the perturbing (Eq.\,(\ref{121030b})) tensors, respectively, expressed in the frame of the central body.

\subsection{Deformation in the frequency domain}
\llabel{deformation_body}

The inertia tensor in the frequency domain can be obtained by performing a Fourier transform
\be
\hat \TI^B(\sigma) = \sqpi
\iinf \TI^B(t) \, \ep^{-\ii \sigma t} \, d t = \hat k_2 (\sigma) \, \hat \TI^B_p (\sigma)
\ , \llabel{200116b} 
\ee
where 
\be
\hat k_2(\sigma) = \sqpi \iinf k_2 (t) \, \ep^{-\ii \sigma t} \, d t 
\llabel{210929e}
\ee
is the second Love number, and we applied the properties of the convolution product.
$\hat k_2(\sigma)$ is a complex number, whose modulus gives the amplitude of the tidal deformation and the argument gives the phase lag between the perturbation and the deformation.
It follows that the deformation in the body frame (Eq.\,(\ref{200116a})) is much easier to compute in the frequency domain (Eq.\,(\ref{200116b})).
However, it is a bit more difficult to compute the deformation in an arbitrary frame.

% Fourier Transform notations:
% https://www.johndcook.com/blog/fourier-theorems/

We let $\SR$ be the rotation matrix that allows us to convert any vector $\vu_B$ in a frame attached to the central body into another frame $\vu$, such that $\vu = \SR \, \vu_B$,
\be
\TI (t) = \SR(t) \, \TI^B(t) \, \SR(t)^T
\ , \quad \mathrm{and} \quad
\TI_p (t) = \SR(t) \, \TI^B_p(t) \, \SR(t)^T   \ .
\llabel{200116c}
\ee

Then
\be
\begin{split}
\hat \TI (\sigma) =&
\sqpi \iinf \TI (t) \, \ep^{-\ii \sigma t} \, d t = 
\sqpi \iinf \SR(t) \, \TI^B(t) \, \SR(t)^T \ep^{-\ii \sigma t} \, d t \\
=& \spi \iinf \iinf k_2(t-\tl) \, \SR(t) \, \TI^B_p(\tl) \, \SR(t)^T \ep^{-\ii \sigma t} \, d \tl d t \\
=& \spi \iinf \iinf k_2(\tu) \, \SR(\tl+\tu) \, \TI^B_p(\tl) \,  \SR(\tl+\tu)^T \ep^{-\ii \sigma (\tl + \tu)} \, d \tl d \tu
\ . \llabel{200116d} 
\end{split}
\ee

Since $\SR(t)$ is a rotation matrix, we note that $\SR(\tl+\tu) = \SR(\tu) \SR(\tl)$, thus
\be
\begin{split}
\hat \TI (\sigma) =& \spi \iinf \iinf k_2(\tu) \, \SR(\tu) \, \TI_p(\tl) \, \SR(\tu)^T \ep^{-\ii \sigma (\tl + \tu)} \, d \tl d \tu \\
=& \sqpi \iinf  k_2(\tu) \, \SR(\tu) \, \hat \TI_p(\sigma) \, \SR(\tu)^T \ep^{-\ii \sigma \tu} \, d \tu
\ . \llabel{200116e} 
\end{split}
\ee

At this stage, we need to explicit $\SR(\tu)$ to proceed.
For simplicity, we adopt the frame $(\vp,\vq,\vs)$, for which we have already obtained the equations of motion (Sect.~\ref{pmp}), and where $\vs$ is the spin axis (Fig.~\ref{frames}).
Since the central body rotates about the $\vs$ axis with velocity $\om$, we have 
\be
\SR(\tu) = \left[\begin{array}{ccc} 
\cos \om \tu   &  -\sin \om \tu   & 0 \\
\sin \om \tu   & \cos \om \tu & 0 \\
0 & 0 & 1
\end{array}\right] 
\ , \llabel{200116f}
\ee
which gives
\be
\begin{split}
& \SR(\tu) \, \hat \TI_p(\sigma) \, \SR(\tu)^T =  \crm 
& \left[\begin{array}{ccccc} 
-\Delta \hat I^\pt \cos 2 \om \tu - \hat I_{12}^\pt \sin 2 \om \tu -\frac12 \hat I_{33}^\pt &,& -\Delta \hat I^\pt \sin 2 \om \tu + \hat I_{12}^\pt \cos 2 \om \tu &,& \hat I_{13}^\pt \cos \om \tu - \hat I_{23}^\pt \sin \om \tu \crm
-\Delta \hat I^\pt \sin 2 \om \tu + \hat I_{12}^\pt \cos 2 \om \tu  &,& \Delta \hat I^\pt \cos 2 \om \tu + \hat I_{12}^\pt \sin 2 \om \tu -\frac12 \hat I_{33}^\pt &,& \hat I_{13}^\pt \sin \om \tu + \hat I_{23}^\pt \cos \om \tu \crm
\hat I_{13}^\pt \cos \om \tu - \hat I_{23}^\pt \sin \om \tu &,& \hat I_{13}^\pt \sin \om \tu + \hat I_{23}^\pt \cos \om \tu &,& \hat I_{33}^\pt
\end{array}\right]
\ , \llabel{200117a}
\end{split}
\ee
where $\Delta \hat I^\pt = \frac12 (\hat I_{22}^\pt - \hat I_{11}^\pt)$.
We thus have, for instance,
\be
\llabel{200117b} 
\begin{split}
\hat I_{13} (\sigma) & = \sqpi
\iinf  k_2(\tu) \left[ \hat I_{13}^\pt (\sigma) \cos \om \tu - \hat I_{23}^\pt (\sigma) \sin \om \tu \right] \ep^{-\ii \sigma \tu} \, d \tu  \\ 
& = \hat I_{13}^\pt (\sigma) \sqpi \iinf  k_2(\tu) \cos \om \tu \, \ep^{-\ii \sigma \tu} \, d \tu -
 \hat I_{23}^\pt (\sigma) \sqpi \iinf  k_2(\tu) \sin \om \tu \, \ep^{-\ii \sigma \tu} \, d \tu  \\
%& = \frac12 \hat I_{13}^\pt (\sigma) \sqpi \iinf  k_2(\tu) \left[ \ep^{\ii \om \tu} + \ep^{-\ii \om \tu}\right] \ep^{-\ii \sigma \tu} \, d \tu +\frac{\ii}{2} \hat I_{23}^\pt (\sigma) \sqpi \iinf  k_2(\tu)\left[ \ep^{\ii \om \tu} - \ep^{-\ii \om \tu}\right] \ep^{-\ii \sigma \tu} \, d \tu  \\
%& = \frac12 \hat I_{13}^\pt (\sigma) \sqpi \iinf  k_2(\tu) \left[ \ep^{-\ii (\sigma-\om) \tu} + \ep^{-\ii (\sigma+\om) \tu}\right]  \, d \tu +\frac{\ii}{2} \hat I_{23}^\pt (\sigma) \sqpi \iinf  k_2(\tu)\left[ \ep^{-\ii (\sigma-\om) \tu} - \ep^{-\ii (\sigma+\om) \tu}\right] \, d \tu  \\
%&= \frac12 \hat I_{13}^\pt (\sigma) [\hat k_2 (\sigma-\om) + \hat k_2 (\sigma+\om)] +\frac{\ii}{2} \hat I_{23}^\pt (\sigma) [\hat k_2 (\sigma-\om) - \hat k_2 (\sigma+\om)] \\
%&= \hat k_2 (\sigma-\om) \frac{\hat I_{13}^\pt (\sigma) + \ii \hat I_{23}^\pt (\sigma)}{2} + \hat k_2 (\sigma+\om) \frac{\hat I_{13}^\pt (\sigma) - i \hat I_{23}^\pt (\sigma)}{2} \\
& = \frac12 \hat k_2 (\sigma-\om) \left[ \hat I_{13}^\pt (\sigma) + \ii \, \hat I_{23}^\pt (\sigma) \right] + \frac12 \hat k_2 (\sigma+\om) \left[ \hat I_{13}^\pt (\sigma) - \ii \, \hat I_{23}^\pt (\sigma) \right] \ .
\end{split}
\ee
Similarly, we obtain for the remaining coefficients of $\hat \TI (\sigma)$:
\be
\hat I_{23} (\sigma) = \frac12 \hat k_2 (\sigma-\om) \left[ \hat I_{23}^\pt (\sigma) - \ii \, \hat I_{13}^\pt (\sigma) \right] + \frac12 \hat k_2 (\sigma+\om) \left[ \hat I_{23}^\pt (\sigma) + \ii \, \hat I_{13}^\pt (\sigma) \right] 
\ , \llabel{200117c} 
\ee
%\be
%\hat I_{23} (\sigma) = \frac12 \left[ \hat k_2 (\sigma+\om) + \hat k_2 (\sigma-\om) \right]  \hat I_{23}^\pt (\sigma) + \ii \, \frac12 \left[ \hat k_2 (\sigma+\om) - \hat k_2 (\sigma-\om) \right] \hat I_{13}^\pt (\sigma)  
%\ , \llabel{200117c} 
%\ee
\be
\hat I_{12} (\sigma) = \frac12 \hat k_2 (\sigma-2\om) \left[ \hat I_{12}^\pt (\sigma) + \ii \, \Delta \hat I^\pt (\sigma) \right] + \frac12 \hat k_2 (\sigma+2\om) \left[ \hat I_{12}^\pt (\sigma) - \ii \, \Delta \hat I^\pt (\sigma) \right] 
\ , \llabel{200117d} 
\ee
\be
\hat I_{33} (\sigma) =\hat k_2 (\sigma) \hat I_{33}^\pt (\sigma)
\ , \llabel{200117f} 
\ee
\be
\hat I_{22} (\sigma) = \Delta \hat I (\sigma) - \frac12 \hat I_{33} (\sigma)
\ , \llabel{200117g} 
\ee
\be
\hat I_{11} (\sigma) = - \Delta \hat I (\sigma) - \frac12 \hat I_{33} (\sigma)
\ , \llabel{200117h} 
\ee
with 
\be
\Delta \hat I (\sigma) = \frac12 \hat k_2 (\sigma-2\om) \left[ \Delta \hat I^\pt (\sigma) - \ii \, \hat I_{12}^\pt (\sigma) \right] + \frac12 \hat k_2 (\sigma+2\om) \left[ \Delta \hat I^\pt (\sigma) + \ii \, \hat I_{12}^\pt (\sigma) \right] 
\ . \llabel{200117e} 
\ee

The coefficients of the inertia tensor $\TI$ (Eq.\,(\ref{121026c})) are now expressed in the frame $(\vp,\vq,\vs)$ and can be directly used in expressions (\ref{170911b}), (\ref{170911c}) and (\ref{151028e}), provided that the coefficients of the perturbing tensor $\TI_\pt$ (Eq.\,(\ref{121030b})) are also given in this frame:
\be
\frac{I_{11}^\pt}{\ms R^2} =  - \frac{\Ms}{\ms} \left(\frac{R}{r }\right)^3 \left( \xii^2 - \frac{1}{3} \right)
\ , \quad
\frac{I_{12}^\pt}{\ms R^2} = - \frac{\Ms}{\ms} \left(\frac{R}{r }\right)^3 \xii \xj
 \ , \llabel{151106a}
\ee
\be
\frac{I_{22}^\pt}{\ms R^2} =  - \frac{\Ms}{\ms} \left(\frac{R}{r }\right)^3 \left( \xj^2 - \frac{1}{3} \right)
\ , \quad
\frac{I_{23}^\pt}{\ms R^2} = - \frac{\Ms}{\ms} \left(\frac{R}{r }\right)^3 \xj  \xk
\ , \llabel{151106b} 
\ee
\be
\frac{I_{33}^\pt}{\ms R^2} =  - \frac{\Ms}{\ms} \left(\frac{R}{r }\right)^3 \left( \xk^2 - \frac{1}{3} \right)
\ , \quad
\frac{I_{13}^\pt}{\ms R^2} = - \frac{\Ms}{\ms} \left(\frac{R}{r }\right)^3 \xii \xk
%\ , \llabel{151106c} 
\ . \llabel{151106f} 
\ee
%\be
%\frac{I_{12}^\pt}{\ms R^2} = - \frac{\Ms}{\ms} \left(\frac{R}{r }\right)^3 \xii \xj
%\ , \llabel{151106d} 
%\ee
%\be
%\frac{I_{13}^\pt}{\ms R^2} = - \frac{\Ms}{\ms} \left(\frac{R}{r }\right)^3 \xii \xk
%\ , \llabel{151106e}
%\ee
%\be
%\frac{I_{23}^\pt}{\ms R^2} = - \frac{\Ms}{\ms} \left(\frac{R}{r }\right)^3 \xj  \xk
%\ . \llabel{151106f} 
%\ee

\subsection{Hansen coefficients}

In order to use the coefficients of the inertia tensor $\hat \TI (\sigma)$ (Eqs.\,(\ref{200117b})$-$(\ref{200117h})) in the equations of motion (Sect.~\ref{pmp}) we need to return to the time domain using an inverse Fourier transform:
\be
\TI (t) = \sqpi \iinf \hat \TI (\sigma) \, \ep^{\ii \sigma t} \, d \sigma 
\ . \llabel{210910a}
\ee
In general, the tidal perturbations introduced by the perturbing tensor $\TI_\pt (t)$ are periodic and thus only a discrete number of frequencies exist, that is, we can express %$\TI_\pt (t)$ and 
$\TI (t)$ as a Fourier series:
\be
%\TI_\pt (t) = \sum_k \hat \TI_\pt (\sigma_k) \, \ep^{\ii \sigma_k t} 
%\ , \quad \mathrm{and} \quad
\TI (t) = \sum_k \hat \TI (\sigma_k) \, \ep^{\ii \sigma_k t} 
\ . \llabel{210910b}
\ee
In the frame $(\vp,\vq,\vs)$, the position of the perturber $m_0$ only depends on its orbital motion (Eq.\,(\ref{210804e})).
Therefore, % in the unperturbed two-body problem, 
the only frequencies are the orbital mean motion, $n$, and its harmonics.
A convenient way of expressing $\TI_p$ is through the Hansen coefficients, $X_k^{\ell,m}$ \citep[e.g.,][]{Hughes_1981}:
\be
\left( \frac{r}{a} \right)^\ell \ep^{\ii m \up} = \sumk X_k^{\ell,m}(e) \, \ep^{\ii k M} 
\ , \llabel{210910c}
\ee
with
\be
X_k^{\ell,m} (e) = \frac{1}{2 \pi} \int_{-\pi}^\pi \left( \frac{r}{a} \right)^\ell \ep^{\ii (m \up-k M)} \, d M
= \frac{1}{\pi} \int_0^\pi \left( \frac{1-e^2}{1+e \cos \up} \right)^\ell \cos (m \up-k M) \, d M
\ . \llabel{210910d}
\ee

%For instance, for the $I_{33}^\pt$ coefficient (Eq.\,(\ref{151106c})), we get from expression (\ref{210804e})
%\be
%\begin{split}
%\frac{I_{33}^\pt}{\ms R^2} = & - \frac{\Ms}{\ms} \left(\frac{R}{a}\right)^3 \left(\frac{r}{a}\right)^{-3} \left[  \left( \cos \up \, \ue \cdot \vs - \sin \up \sin \theta \, \ue \cdot \vp \right)^2 - \frac{1}{3} \right] \\
%= & - \frac{\Ms}{\ms} \left(\frac{R}{a}\right)^3 \sumk \left(\frac{r}{a}\right)^{-3} \left[  \left( \cos \up \, \ue \cdot \vs - \sin \up \sin \theta \, \ue \cdot \vp \right)^2 - \frac{1}{3} \right] \ep^{\ii k M} 
%\ , \llabel{210910e} 
%\end{split}
%\ee
%\be
%\begin{split}
%\frac{I_{33}^\pt}{\ms R^2} = & - \frac{\Ms}{\ms} \left(\frac{R}{a}\right)^3 \left(\frac{r}{a}\right)^{-3} \left[ \sin^2 \theta \sin^2 (\vpi+\up) - \frac{1}{3} \right] \\
%= & - \frac{\Ms}{\ms} \left(\frac{R}{a}\right)^3 \left(\frac{r}{a}\right)^{-3} \left[ \frac14 \sin^2 \theta \left(2 -  \ep^{\ii 2 (\vpi+\up)} - \ep^{-\ii 2 (\vpi+\up)}  \right) - \frac{1}{3} \right] \\
%= & - \frac{\Ms}{\ms} \left(\frac{R}{a}\right)^3 \sumk \left[ \frac14 \sin^2 \theta \left(2 X_k^{-3,0} -  \ep^{\ii 2 \vpi} X_k^{-3,2} - \ep^{-\ii 2 \vpi} X_k^{-3,-2}  \right) - \frac{1}{3} X_k^{-3,0} \right] \ep^{\ii k M} 
%\ , \llabel{210910f} 
%\end{split}
%\ee

For instance, for the $I_{23}^\pt$ coefficient (Eq.\,(\ref{151106b})), we get from expression (\ref{210804e})
\be
\begin{split}
\frac{I_{23}^\pt}{\ms R^2} %&= - \frac{\Ms}{\ms} \left(\frac{R}{r }\right)^3 \xj  \xk \crm
& = \frac{\Ms}{\ms} \left(\frac{R}{r }\right)^3  \sin \theta \cos \theta \sin^2 (\vpi+\up) \crm
%& = \frac12 \frac{\Ms}{\ms} \left(\frac{R}{r }\right)^3  \sin \theta \cos \theta  \left( 1 - \cos  (2 \vpi+ 2\up) \right) \crm
 & = \frac14 \frac{\Ms}{\ms} \left(\frac{R}{a}\right)^3  \sin \theta \cos \theta  \left(\frac{a}{r}\right)^3 \left(2 -  \ep^{\ii 2 (\vpi+\up)} - \ep^{-\ii 2 (\vpi+\up)}  \right) \crm
& =  \frac14 \frac{\Ms}{\ms} \left(\frac{R}{a}\right)^3 \sin \theta \, \sumk \cT \left(2 X_k^{-3,0} -  \ep^{\ii 2 \vpi} X_k^{-3,2} - \ep^{-\ii 2 \vpi} X_k^{-3,-2}  \right)  \ep^{\ii k M}
\ , \llabel{211001e} 
\end{split}
\ee
where $\cT = \cos \theta = \vk \cdot \vs $, and (Eq.\,(\ref{210804d}))
\be
\ep^{\ii \varpi} = \ue \cdot \vp - \ii \frac{\ue \cdot \vs }{\sin \theta} 
%\ , \quad \mathrm{and} \quad
\quad \Rightarrow \quad
\ep^{\ii 2 \varpi} = (\ue \cdot \vp)^2 - \left( \frac{\ue \cdot \vs }{\sin \theta} \right)^2 - 2 \ii \, (\ue \cdot \vp) \left(\frac{\ue \cdot \vs }{\sin \theta}\right)
\ . \llabel{210915a}
\ee
%\be
%\sin^2 \theta \, \ep^{\ii 2 \varpi} = \sin^2 \theta \, (\ue \cdot \vp)^2 - ( \ue \cdot \vs )^2 - 2 \ii \sin \theta \, (\ue \cdot \vp) (\ue \cdot \vs )
%\ . \llabel{210915a}
%\ee
Similarly, for the $I_{13}^\pt$ coefficient (Eq.\,(\ref{151106f})), we get
\be
\frac{I_{13}^\pt}{\ms R^2} % = - \frac{\Ms}{\ms} \left(\frac{R}{r }\right)^3 \xii \xk \crm
%= \frac{\Ms}{\ms} \left(\frac{R}{r }\right)^3  \frac{\sin \theta}{2}  \sin (2 \vpi+ 2\up) \crm
%= \frac{\Ms}{\ms} \left(\frac{R}{r }\right)^3 \frac{\sin \theta}{4 \ii} \left(  \ep^{\ii (2 \vpi+ 2\up)} -  \ep^{-\ii (2 \vpi+ 2\up)}  \right) \crm
=  \frac{1}{4 \ii} \frac{\Ms}{\ms} \left(\frac{R}{a}\right)^3 \sin \theta \, \sumk  \left(  \ep^{\ii 2 \vpi} X_k^{-3,2} -  \ep^{-\ii 2 \vpi} X_k^{-3,-2} \right) \ep^{\ii k M}
\ . \llabel{211001d} 
\ee
%Then, for the $I_{33}$ coefficient we have (Eqs.\,(\ref{200117f}) and (\ref{210910b}))
%\be
%\frac{I_{33}}{\ms R^2}  = - \frac{\Ms}{\ms} \left(\frac{R}{a}\right)^3 \sumk  \hat k_2 (k n) \left[ \frac14 \sin^2 \theta \left(2 X_k^{-3,0} -  \ep^{\ii 2 \vpi} X_k^{-3,2} - \ep^{-\ii 2 \vpi} X_k^{-3,-2}  \right) - \frac{1}{3} X_k^{-3,0} \right] \ep^{\ii k M}
%\ . \llabel{210910g} 
%\ee
Then, for the $I_{23}$ coefficient we finally have (Eqs.\,(\ref{200117c}) and (\ref{210910b}))
%\begin{eqnarray}  
%I_{23}  = \frac{\Ms R^5}{8 a^3} \sin \theta \sumk & \bigg\{ & \cos \theta \,  \left[ \hat k_2 (k n+\om) + \hat k_2 (k n-\om) \right] \left(2 X_k^{-3,0} -  \ep^{\ii 2 \vpi} X_k^{-3,2} - \ep^{-\ii 2 \vpi} X_k^{-3,-2} \right) \crm
%& +& \left[ \hat k_2 (k n+\om) - \hat k_2 (k n-\om) \right]  \left(  \ep^{\ii 2 \vpi} X_k^{-3,2} -  \ep^{-\ii 2 \vpi} X_k^{-3,-2} \right) \bigg\} \, \ep^{\ii k M} 
%\ . \llabel{211001f} 
%\end{eqnarray}
\begin{eqnarray}
\llabel{211001f} 
I_{23}  = \frac{\Ms R^5}{8 a^3} \sin \theta \sumk & \bigg[ &
\hat k_2 (kn-\om) \left(2 \cT X_k^{-3,0} - (1+\cT)  \ep^{\ii 2 \vpi} X_k^{-3,2} + (1-\cT) \ep^{-\ii 2 \vpi} X_k^{-3,-2}  \right) \\
& +& \hat k_2 (kn+\om) \left(2 \cT X_k^{-3,0} + (1- \cT)  \ep^{\ii 2 \vpi} X_k^{-3,2} - (1+\cT) \ep^{-\ii 2 \vpi} X_k^{-3,-2}  \right) \bigg] \, \ep^{\ii k M} \nonumber \ . 
\end{eqnarray}
%os vetores $\ve$ e $\vp$ tambem se movem, mas lentamente e so devido aos efeitos de mare, pelo que sao considerados constantes aqui

Since the coefficients of the $\TI_\pt$ tensor are all $I_{ij}^\pt \propto r^{-3}$ (Eqs.\,(\ref{151106a})$-$(\ref{151106f})), in the expression of the inertia tensor $\TI$ can only appear Hansen coefficients of the kind $X_k^{-3,m}$.
The same occurs with the torque (Eq.\,(\ref{151028e})).
The force (Eq.\,(\ref{170911d})) and the derivative of the Laplace vector (Eq.\,(\ref{210805c})) are proportional to $r^{-4}$ and we thus expect terms in $X_k^{-4,m}$.
However, it is possible to provide the entire set of equations of motion only in terms of the coefficients $X_k^{-3,0}$, $X_k^{-3,1}$, and $X_k^{-3,2}$ (Table~\ref{tabHansen}), by using some recorrence properties of the Hansen coefficients (see appendix~\ref{hcr}).

\begin{table*}
\begin{center}
\begin{tabular}{|r|c|c|c| } \hline 
$k$ & $X_k^{-3,0} (e) $ & $X_k^{-3,1} (e) $ & $X_k^{-3,2} (e) $ \\ \hline
$-6$ & $\frac{3167}{320} e^6$ & $-$ & $-$ \\
$-5$ & $\frac{1773}{256} e^5$ & $\frac{16289}{9216} e^6$ & $-$ \\
$-4$ & $\frac{77}{16} e^4 + \frac{129}{160} e^6$ & $\frac{643}{480} e^5$ & $\frac{4}{45} e^6$ \\
$-3$ & $\frac{53}{16} e^3 + \frac{393}{256} e^5$ & $\frac{131}{128} e^4 + \frac{237}{256} e^6$ & $\frac{81}{1280} e^5$ \\
$-2$ & $\frac{9}{4} e^2 + \frac{7}{4} e^4 + \frac{141}{64} e^6$ & $\frac{19}{24} e^3 + \frac{43}{48} e^5$ & $\frac{1}{24} e^4 + \frac{7}{240} e^6$ \\
$-1$ & $\frac{3}{2} e + \frac{27}{16} e^3 + \frac{261}{128} e^5$ & $\frac{5}{8} e^2 + \frac{5}{6} e^4 + \frac{3103}{3072} e^6$ & $\frac{1}{48} e^3 + \frac{11}{768} e^5$ \\
$0$ & $1 + \frac{3}{2} e^2 + \frac{15}{8} e^4 + \frac{35}{16} e^6$ & $\frac{1}{2} e + \frac{3}{4} e^3 + \frac{15}{16} e^5$ & $-$ \\
$1$ & $\frac{3}{2} e + \frac{27}{16} e^3 + \frac{261}{128} e^5$ & $1 + \frac{1}{2} e^2 + \frac{55}{64} e^4 + \frac{1177}{1152} e^6$ & $- \frac{1}{2} e + \frac{1}{16} e^3 - \frac{5}{384} e^5$ \\
$2$ & $\frac{9}{4} e^2 + \frac{7}{4} e^4 + \frac{141}{64} e^6$ & $\frac{5}{2} e - \frac{1}{8} e^3 + \frac{103}{96} e^5$ & $1 - \frac{5}{2} e^2 + \frac{13}{16} e^4 - \frac{35}{288} e^6$ \\
$3$ & $\frac{53}{16} e^3 + \frac{393}{256} e^5$ & $\frac{39}{8} e^2 - 2 \, e^4 + \frac{1803}{1024} e^6$ & $\frac{7}{2} e - \frac{123}{16} e^3 + \frac{489}{128} e^5$ \\
$4$ & $\frac{77}{16} e^4 + \frac{129}{160} e^6$ & $\frac{103}{12} e^3 - \frac{593}{96} e^5$ & $\frac{17}{2} e^2 - \frac{115}{6} e^4 + \frac{601}{48} e^6$ \\
$5$ & $\frac{1773}{256} e^5$ & $\frac{5485}{384} e^4 - \frac{11053}{768} e^6$ & $\frac{845}{48} e^3 - \frac{32525}{768} e^5$ \\
$6$ & $\frac{3167}{320} e^6$ & $\frac{3669}{160} e^5$ & $\frac{533}{16} e^4 - \frac{13827}{160} e^6$ \\
$7$ & $-$ & $\frac{330911}{9216} e^6$ & $\frac{228347}{3840} e^5$ \\
$8$ & $-$ & $-$ & $\frac{73369}{720} e^6$ \\ \hline
\end{tabular} 
\end{center}
\caption{Hansen coefficients up to $ e^6 $. The exact expression of these coefficients is given by expression (\ref{210910d}). \llabel{tabHansen} }  
\end{table*}

\subsection{Tidal models}

\llabel{tidalmodels}

The tidal deformation of the central body is completely described by expression (\ref{210910b}), where $\hat \TI (\sigma)$ is obtained from the Love number $\hat k_2 (\sigma)$ combined with the perturbing tensor $\hat \TI_\pt (\sigma)$ (Eqs.\,(\ref{200117b}$)-($\ref{200117h})).
While the perturbing tensor is well determined, as it depends only on the position of the perturbing body (Eqs.\,(\ref{151106a})$-$(\ref{151106f})), the Love number is subject to large uncertainties, as it depends on the internal structure of the central body.
Therefore, in order to compute $\hat k_2 (\sigma)$ one needs to adopt some rheological model for the deformation.
A large variety of models exist, but the most commonly used are the constant$-Q$ \citep[e.g.,][]{Munk_MacDonald_1960}, the linear model \citep[e.g.,][]{Mignard_1979}, 
the Maxwell model \citep[e.g.,][]{Correia_etal_2014},
and the Andrade model \citep[e.g.,][]{Efroimsky_2012b}.
Some models appear to be better suited to certain situations, but there is no model that is globally accepted.
However, viscoelastic rheologies are usually more realistic as they are able to reproduce the main features of tidal dissipation \citep[e.g.][]{Remus_etal_2012a}.
For a review of the main viscoelastic models see \citet{Renaud_Henning_2018}.
The Love number is a complex number, and so it can be decomposed in its real and imaginary parts as
\be
\hat k_2 (\sigma) = a (\sigma) - \ii \, b (\sigma) 
\ . \llabel{210924d}
\ee
This partition is very useful when we write the equations of motion (see Sects.~\ref{singav} and~\ref{doubav}), because the imaginary part characterises the material's viscous phase lag and is thus directly related to the amount of energy dissipated by tides.
This is why we write expression (\ref{210924d}) with a minus for the imaginary part, the deformation lags behind the perturbation and therefore the imaginary part is always negative.
\bfx{In addition, from expression (\ref{210929e}) we have $\hat k_2^* (\sigma) = \hat k_2 (-\sigma)$, and so $a (\sigma) =  a (-\sigma) $ is always an even function and $b (\sigma) = - b (-\sigma)$ is always an odd function.}
Here, we provide the expressions of $a (\sigma)$ and $b (\sigma)$ for the more frequently used tidal models.

%The Love number is a complex number, and so it can be decomposed in its real and imaginary parts as
%\be
%\hat k_2 (\sigma) =  |\hat k_2 (\sigma)| \ep^{-\ii \epsilon(\sigma)} = a (\sigma) - \ii \, b (\sigma) \ , \llabel{210924d}
%\ee
%\bfx{where $\epsilon(\sigma)$ is the phase lag.}
%This partition is very useful when we write the equations of motion (see sections~\ref{singav} and~\ref{doubav}), because the imaginary part is directly related to the amount of energy dissipated by tides.
%This is why we write expression (\ref{210924d}) with a minus for the imaginary part, the deformation lags behind the perturbation and therefore the imaginary part is always negative \citep[e.g.,][]{Efroimsky_2012b}.
%\bfx{Note also that, for symmetry reasons, $|\hat k_2 (-\sigma)| = |\hat k_2 (\sigma)|$ and $\epsilon(-\sigma) = - \epsilon(\sigma)$.
%As a consequence, $a (\sigma) = |\hat k_2 (\sigma)| \cos \epsilon(\sigma)$ is always an even function and $b (\sigma) = |\hat k_2 (\sigma)| \sin \epsilon(\sigma)$ is always an odd function}. 
%Here, we provide the expressions of $a (\sigma)$ and $b (\sigma)$ for the more frequently used tidal models.

\subsubsection{Constant$-Q$ model}

A commonly used dimensionless measure of the tidal dissipation is the quality factor % or $Q-$factor, 
\be
Q^{-1} = \frac{1}{2 \pi E_0} \oint \dot E \, d t 
\ , \llabel{131004b}
\ee
where the line integral over $\dot E$ is the energy dissipated during one period of tidal stress, and $E_0$ is the peak energy stored in the system during the same period.
In general, $Q$ is a function of the frequency, and can be related to the Love number through \citep[e.g.,][]{Henning_etal_2009}
\be
b (\sigma) = \frac{|\hat k_2(\sigma)|}{Q (\sigma)}
\ . \llabel{211001a}
\ee
The constant$-Q$ model assumes that both the quality factor and the norm of the Love number are constant for all frequencies, that is, $Q(\sigma) = Q(0) = Q_0$ and $|\hat k_2(\sigma)| = |\hat k_2(0)| = \kf$.
The quantity $\kf$ is the fluid Love number, which corresponds to the maximal deformation resulting from a permanent perturbation and depends only on the internal structure of the central body. 
For instance, for a homogenous sphere, we have $\kf = 3/2$.
Since $b (\sigma)$ is an odd function we have
\be
a(\sigma) = \kf 
\ , \quad \mathrm{and} \quad
b(\sigma) = \frac{\kf}{Q_0} \, \mathrm{sign} (\sigma)
% = \frac{\kf}{Q_0} \frac{\sigma}{|\sigma|}
\ . \llabel{211001b}
\ee
\bfx{The constant$-Q$ model is appropriate for short evolution timescales, where the orbital and the spin frequencies do not change dramatically.}

\subsubsection{Linear or weak friction model}

The linear or weak friction model %, also known as the viscous linear model, 
assumes that the time delay, $\Delta t$, between the maximal deformation and the perturbation is constant and small \citep{Singer_1968, Alexander_1973, Mignard_1979}.
This is equivalent to assume that the Love distribution is a Dirac distribution \citep[see][]{Boue_etal_2019p}
\be
k_2 (t) = \kf \, \delta (t - \Delta t) 
\ . \llabel{210929d}
\ee
As a result, we have (Eq.\,(\ref{210929e}))
\be
\hat k_2(\sigma) %= \kf \sqpi \iinf \delta (t - \tv)  \, \ep^{-\ii \sigma t} \, d t 
= \kf \, \ep^{- \ii \sigma \Delta t } 
\approx \kf \, (1 - \ii \sigma \Delta t)
\ , \llabel{210929f}
\ee
that is
\be
a(\sigma) = \kf 
\ , \quad \mathrm{and} \quad
b(\sigma) = \kf \, \sigma \Delta t 
\ . \llabel{210929c}
\ee
The time delay can be related to the Newton fluid relaxation time, $\Delta t = \tv$, with
\be
\tv % = \frac{19 \eta}{2 g \rho R}
= \frac{38 \pi \eta R^4}{3 \Gc \ms^2}
\ , \llabel{210929b}
\ee
where $\eta$ is the viscosity of the fluid.
%, $g$ is the mean gravity and $\rho$ is the mean density. 
\bfx{The linear model is appropriate for bodies with small viscosities. %, in which case the deformations are small and lags can be assumed proportional to the frequencies.
It is also a good approximation of other tidal models when $\sigma \approx 0 $.}

\subsubsection{Maxwell model}

A material is called a Maxwell solid when it responds to stresses like a massless, damped harmonic oscillator.
It is characterised by a rigidity $\mu$ (or shear modulus), and by a viscosity $\eta$. 
It is one of the simplest viscoelastic models, where the material behaves like an elastic solid over short time scales ($\eta \rightarrow \infty$), but flows like a fluid over long periods of time ($\mu \rightarrow \infty$).
The Love number for the Maxwell model is given by \citep[e.g.][]{Darwin_1908}:
\be
\hat k_2 (\sigma) = \kf \, \frac{1 + \ii \sigma \te}{1 + \ii \sigma \tau}
\ , \llabel{210929a}
\ee
where $\te = \eta / \mu$ is the elastic or Maxwell relaxation time and $\tau = \te + \tv$.
We thus have
\be
a(\sigma) = \kf \, \frac{1 + \sigma^2 \te \tau}{1 + (\sigma \tau)^2}
\ , \quad \mathrm{and} \quad
b(\sigma) = \kf \, \frac{\sigma \tv}{1 + (\sigma \tau)^2}
\ . \llabel{210929z}
\ee
\bfx{The Maxwell model is appropriate for bodies with a rocky nature. 
For small tidal frequencies $(\sigma \tau \ll 1)$ it is similar to the linear model, but for high tidal frequencies $(\sigma \tau \gg 1)$ it becomes inversely proportional to the tidal frequency, $\sigma$. 
This feature gives rise to the appearance of non-synchronous spin-orbit resonances in this regime \citep[e.g.,][]{Correia_etal_2014}.}

\subsubsection{Andrade model}

The Andrade model is also a viscoelastic model, but more complex than the Maxwell one.
It is derived from laboratory measurements on the response of materials to stress.
It can be represented by three mechanical elements combined in series, a spring, a dashpot and a spring-pot \citep[e.g.,][]{Andrade_1910, BenJazia_etal_2014, Gevorgyan_etal_2020}.
The Love number for the Andrade model is given by \citep[e.g.,][]{Efroimsky_2012b}:
\be
\hat k_2 (\sigma) = \frac{\kf}{1 + \hat \mu (\sigma)}
\ , \llabel{210930a}
\ee
where % $\tau_M = \te$
\be
\hat \mu (\sigma) % = A_2 \mu(\sigma) / \mu 
%=  \frac{\tv / \te}{1 - \ii (\sigma \te)^{-1} + (\ii \sigma \ta)^{-\alpha} \Gamma (1+\alpha)  }
=  \frac{\tv}{\te} \left[1 - \ii (\sigma \te)^{-1} + (\ii \sigma \ta)^{-\alpha} \Gamma (1+\alpha) \right]^{-1}
\ , \llabel{210930b}
\ee
is the effective rigidity.
The parameter $\alpha$ is an empirical adjustable parameter whose value depends on the material, we usually have $ 0.2 \le \alpha \le 0.4$.
The quantity $\ta$ is the timescale associated with the Andrade creep and may be termed as the ``Andrade'' or the ``anelastic'' time.
Then, % after some algebra we get
\be
a(\sigma) = \kf \left[ 1 - \frac{{\cal A} (\sigma) \, \sigma \tv }{{\cal A} (\sigma)^2 + {\cal B} (\sigma)^2} \right]
\ , \quad \mathrm{and} \quad
b(\sigma) = \kf \, \frac{{\cal B} (\sigma) \, \sigma \tv}{{\cal A} (\sigma)^2 + {\cal B} (\sigma)^2}
\ , \llabel{210930c}
\ee
with
\be
{\cal A} (\sigma) = (\sigma \tau) \left[ 1 + |\sigma \tau|^{-\alpha} \left(\frac{\te}{\tau}\right) \left(\frac{\tau}{\ta}\right)^\alpha \cos \left(\frac{\alpha \pi}{2}\right) \Gamma (1+\alpha) \right]
\ , \llabel{210930d}
\ee
and
\be
{\cal B} (\sigma) = 1 + |\sigma \tau|^{1-\alpha} \left(\frac{\te}{\tau}\right) \left(\frac{\tau}{\ta}\right)^\alpha \sin \left(\frac{\alpha \pi}{2}\right) \Gamma (1+\alpha)
\ . \llabel{210930e}
\ee
%Note that for $(\sigma \tau) \ll 1$ both the Andrade and Maxwell models resume to the linear model.
%Note also that for $\alpha = 1$ we retrieve the expressions for the Maxwell model.
\bfx{The Andrade model is also appropriate for bodies with a rocky nature, which includes the effect of transient creep response.
As a result, for high tidal frequencies $(\sigma \tau \gg 1)$ it dissipates more energy than in the case of the Maxwell model \citep[see discussions in][]{Renaud_Henning_2018}.}

\section{Average over the mean anomaly}
\llabel{singav}

In general, tidal effects slowly modify the spin and the orbit of the central body, in a timescale much longer than the orbital period of the system.
Therefore, we can average the equations of motion (Sect.~\ref{pmp}) over the mean anomaly and obtain the equations for the secular evolution of the system due to tidal effects.

In order to proceed with this calculation, we first need to expand the tidal torque (Eq.\,(\ref{151028e})) and the derivative of the Laplace vector (Eq.\,(\ref{210805c})) in Hansen coefficients (Eq.\,(\ref{210910c})). 
For instance, for the last term in the $\vs$ component of the tidal torque we have (Eq.\,(\ref{210804e}))
\be
\begin{split}
%\vT \cdot \vs [4] % =
 - \frac{3 \Gc \Ms}{r^3} I_{23} \xii \xk 
& = \frac{3 \Gc \Ms}{2 r^3} I_{23}  \sin \theta \sin (2 \vpi+2\up) \crm
%& = \frac{3 \Gc \Ms}{4 \ii r^3} I_{23}  \sin \theta \left( \ep^{\ii (2 \vpi+2\up)} - \ep^{-\ii (2 \vpi+2\up)} \right) \crm
& = \frac{3 \Gc \Ms}{4 \ii a^3} I_{23}  \sin \theta \, \sumkl \left( \ep^{\ii 2 \vpi} X_\kl^{-3,2} - \ep^{-\ii 2 \vpi} X_\kl^{-3,-2}  \right) \ep^{\ii \kl M}
\ . \llabel{211001c}
\end{split}
\ee

Then, we replace the expression of $I_{23}$ also expanded in Hansen coefficients (Eq.\,(\ref{211001f})), and average over the mean anomaly, $M$, which is equivalent to retain only the terms with $\kl = -k$:
%\begin{eqnarray}
%\left\langle - \frac{3 \Gc \Ms}{r^3} I_{23} \xii \xk \right\rangle_M 
%& = & \frac{3 \Gc \Ms^2 R^5}{32 \ii a^6}  \sin^2 \theta \, \sumk \left( \ep^{\ii 2 \vpi} X_{-k}^{-3,2} - \ep^{-\ii 2 \vpi} X_{-k}^{-3,-2}  \right)  \crm
%& & \bigg\{ \cos \theta \left[ \hat k_2 (k n+\om) + \hat k_2 (k n-\om) \right] \left(2 X_k^{-3,0} -  \ep^{\ii 2 \vpi} X_k^{-3,2} - \ep^{-\ii 2 \vpi} X_k^{-3,-2}  \right) \crm
%& & + \left[ \hat k_2 (k n+\om) - \hat k_2 (k n-\om) \right]  \left(  \ep^{\ii 2 \vpi} X_k^{-3,2} -  \ep^{-\ii 2 \vpi} X_k^{-3,-2} \right) \bigg\}
%\ . \llabel{211006a}
%\end{eqnarray}
\be
 \llabel{211006a}
\begin{split}
\left\langle - \frac{3 \Gc \Ms}{r^3} I_{23} \xii \xk \right\rangle_M 
& =  \frac{3 \Gc \Ms^2 R^5}{32 \ii a^6}  \sin^2 \theta \sumk \left( \ep^{\ii 2 \vpi} X_{-k}^{-3,2} - \ep^{-\ii 2 \vpi} X_{-k}^{-3,-2}  \right) \times \\
&  \bigg[ \hat k_2 (kn-\om) \left(2 \cT X_k^{-3,0} - (1+\cT)  \ep^{\ii 2 \vpi} X_k^{-3,2} + (1-\cT) \ep^{-\ii 2 \vpi} X_k^{-3,-2}  \right)  \\
&  + \hat k_2 (kn+\om) \left(2 \cT X_k^{-3,0} + (1- \cT)  \ep^{\ii 2 \vpi} X_k^{-3,2} - (1+\cT) \ep^{-\ii 2 \vpi} X_k^{-3,-2}  \right) \bigg] 
\ .
\end{split}
\ee

Finally, we decompose the Love number in its real and imaginary parts (Eq.\,(\ref{210924d})), make use of their parity properties, and use the simplification $X_{-k}^{-3,m} = X_k^{-3,-m}$ (Eq.\,(\ref{210915b})) to write
%\begin{eqnarray}
%\left\langle - \frac{3 \Gc \Ms}{r^3} I_{23} \xii \xk \right\rangle_M 
%& = & - \frac{3 \Gc \Ms^2 R^5}{32 a^6}  \sin^2 \theta \, \sumk \left( \ep^{\ii 2 \vpi} X_k^{-3,-2} - \ep^{-\ii 2 \vpi} X_k^{-3,2}  \right)  \\
%& & \bigg[  \ii a (\om-kn) \left(2 \cT X_k^{-3,0} - (1+\cT)  \ep^{\ii 2 \vpi} X_k^{-3,2} + (1-\cT) \ep^{-\ii 2 \vpi} X_k^{-3,-2}  \right) \nonumber \\
%& & +  \ii a (\om+kn)  \left(2 \cT X_k^{-3,0} + (1- \cT)  \ep^{\ii 2 \vpi} X_k^{-3,2} - (1+\cT) \ep^{-\ii 2 \vpi} X_k^{-3,-2}  \right)  \nonumber \\
%& & - b (\om-kn)  \left(2 \cT X_k^{-3,0} - (1+\cT)  \ep^{\ii 2 \vpi} X_k^{-3,2} + (1-\cT) \ep^{-\ii 2 \vpi} X_k^{-3,-2}  \right) \nonumber \\
%& & +  b (\om+kn) \left(2 \cT X_k^{-3,0} + (1- \cT)  \ep^{\ii 2 \vpi} X_k^{-3,2} - (1+\cT) \ep^{-\ii 2 \vpi} X_k^{-3,-2}  \right) \bigg] \nonumber \ .
%\end{eqnarray}  
%\begin{eqnarray}
%\left\langle - \frac{3 \Gc \Ms}{r^3} I_{23} \xii \xk \right\rangle_M 
%& = &  \frac{3 \Gc \Ms^2 R^5}{16 a^6}  \sin^2 \theta \, \sumk   \\
%& & \bigg[  \ii a (\om-kn) \cT \left(X_k^{-3,2} X_k^{-3,-2}  ( \ep^{\ii 4 \vpi} - \ep^{-\ii 4 \vpi} ) - X_k^{-3,0} ( X_k^{-3,2} + X_k^{-3,-2} )( \ep^{\ii 2 \vpi} - \ep^{-\ii 2 \vpi} )\right) \nonumber \\
%& & + b (\om-kn)  \left( (1+x) (X_k^{-3,2})^2 + (1-x) (X_k^{-3,-2})^2 - X_k^{-3,2} X_k^{-3,-2}  ( \ep^{\ii 4 \vpi} + \ep^{-\ii 4 \vpi} ) \right. \nonumber \\
%&& + \left. \cT X_k^{-3,0} ( X_k^{-3,2} - X_k^{-3,-2} )( \ep^{\ii 2 \vpi} + \ep^{-\ii 2 \vpi} )   \right) \bigg] \nonumber \ .
%\end{eqnarray}
\be
\llabel{211006b}
\begin{split}
\left\langle - \frac{3 \Gc \Ms}{r^3} I_{23} \xii \xk \right\rangle_M 
& =   \frac{3 \Gc \Ms^2 R^5}{16 a^6}  \sin^2 \theta \sumk \Bigg\{  
b (\om-kn) \bigg[ 2 \cT \cos 2 \vpi \, X_k^{-3,0} \left( X_k^{-3,-2} - X_k^{-3,2} \right)   \\
 & - 2 \cos 4 \vpi  \, X_k^{-3,-2} X_k^{-3,2}  + (1-x) \left(X_k^{-3,-2}\right)^2 +  (1+x) \left(X_k^{-3,2}\right)^2  \bigg]  \\
& + 2 \cT \, a (\om-kn)  \bigg[ \sin 2 \vpi \, X_k^{-3,0} \left( X_k^{-3,-2} + X_k^{-3,2} \right)  -  \sin 4 \vpi  \, X_k^{-3,-2} X_k^{-3,2}  \bigg] \Bigg\} \ ,
\end{split}
\ee
This last arrangement of the Love number is very useful, because terms in $a(\sigma)$ correspond to conservative contributions to the equations of motion, while the terms in $b(\sigma)$ are responsible for the dissipation and consequent tidal evolution.
In addition, we are able to combine terms in $\hat k_2 (\om \pm kn)$ in a single term $\hat k_2 (\om-kn)$.

\subsection{Tidal torque}

The tidal torque is responsible for the variations that occur in the angular momenta (Eqs.\,(\ref{150626a}) and (\ref{210805b})).
In the reference frame $(\vp, \vq, \vs)$ its average value is obtained from expression (\ref{151028e}) as
\be
\big\langle \vT \big\rangle_M = T_p \, \vp + T_q \, \vq + T_s \, \vs 
\ . \llabel{211017a}
\ee
However, we verify there exist some interesting symmetries in the expression of the average torque, which simplifies if we introduce two additional projections 

\be
\big\langle \vT \big\rangle_M   =  \hat T_p \, \vp + \hat T_q \, \vq + \hat T_s \, \vs + T_4 \, \ue + T_5 \, \vs \times \ue 
\ , \llabel{211017b}
\ee
with
\be
\hat T_p % =  %T_p - T_4 \, \ue \cdot \vp - T_5 \, (\vs \times \ue) \cdot \vp 
= T_p - T_4 \, \ue \cdot \vp + T_5 \, \ue \cdot \vq  \ , \quad 
\hat T_q % = % T_q - T_4 \, \ue \cdot \vq - T_5 \, (\vs \times \ue) \cdot \vq 
= T_q - T_5 \, \ue \cdot \vp - T_4 \, \ue \cdot \vq  \ , \quad 
\hat T_s  = T_s - T_4 \, \ue \cdot \vs 
\ , \llabel{211017c}
\ee
or,
%\be
%\ue \cdot \vp = \vs \cdot (\ue \times \vk) / \sin \theta = - \eks / \sin \theta
%\ , \quad \mathrm{and} \quad
%\ue \cdot \vq = - (\ue \cdot \vs) \cos \theta / \sin \theta = -  \cT \es/ \sin \theta
%\ee
\be
\hat T_p  = T_p + \frac{T_4 \eks - T_5 \cT \es}{\sin \theta}  \ , \quad 
\hat T_q  = T_q + \frac{T_4 \cT \es + T_5 \eks}{\sin \theta}  \ , \quad 
\hat T_s  = T_s - T_4 \es 
\ , \llabel{211017c}
\ee
where we used the following notations:
\be
\cT = \vk \cdot \vs = \cos \theta
 \ , \quad 
\es = \ue \cdot \vs = - \sin \theta \sin \vpi 
 \ , \quad \mathrm{and} \quad
\eks = (\vk \times \ue) \cdot \vs %= - \sin \theta \, \ue \cdot \vp 
= - \sin \theta \cos \vpi 
\ . \llabel{211026a}
\ee

This rearrangement of the projections removes apparent singularities for $e=0$, since all terms in $\cos \vpi$ and $\sin \vpi$ are transferred to the Laplace vector $\ve$ (Eq.\,(\ref{210915a})).
Similarly, in order to remove the apparent singularities for $\theta=0$ we finally write
%\be
%\overline \vT = \frac{\hat T_q}{\sin \theta} \, \vk + \left( \hat T_s - \cos \theta \frac{\hat T_q}{\sin \theta} \right) \vs + \frac{\hat T_p}{\sin \theta} \, \vk \times \vs + T_4 \, \ue + T_5 \, \vs \times \ue
%\ , \llabel{211017d}
%\ee
\be
\big\langle \vT \big\rangle_M   = T_1 \, \vk + T_2 \, \vs + T_3 \, \vk \times \vs + T_4 \, \ue + T_5 \, \vs \times \ue 
\ , \llabel{211017d}
\ee
with
\be
T_1 =  \frac{\hat T_q}{\sin \theta}  \ , \quad
T_2 = \left( \hat T_s - \cos \theta \frac{\hat T_q}{\sin \theta} \right) \ , \quad
T_3 = \frac{\hat T_p}{\sin \theta} 
\ .  \llabel{211017e}
\ee
The expressions of the coefficients $T_1, ... , T_5 $ do not present any singularities and can be obtained solely from the angular momentum and Laplace unit vectors as
%\be \hat T_p =  {\sf Txc[1]} \ ; \quad \hat T_q =  {\sf Txc[2]} \ ; \quad \hat T_s =  {\sf Txc[3]} \ee
%\be T_4 =  {\sf T4c} \ ; \quad T_5 =  {\sf T5c} \ ; \quad \hat T_s - \cos \theta \frac{\hat T_q}{\sin \theta} =  {\sf T2c} \ee
%
\be
\small
\begin{split}
\llabel{211027t1}
T_1 &=
%\frac{\sf Txc[2]}{\sT} = 
- \At \sumk \Bigg\{ \dfrac{3}{32} \, \bz  \bigg[
   3 \left(1-\cT^2\right) \left(\left(\Xtdm\right)^2 - \left(\Xtdp\right)^2\right)    \\
& - 2 \left(1-3 \cT^2-2 \es^2\right) \Xtz \left(\Xtdm - \Xtdp\right)
 \bigg]  \\
&  + \dfrac{3}{16} \, \bu  \bigg[   
   4 \cT^3 \left(\Xtz\right)^2
 + \left(1-\cT\right)^2 \left(2+\cT\right) \left(\Xtdm\right)^2
 - \left(1+\cT\right)^2 \left(2-\cT\right) \left(\Xtdp\right)^2   \\&
 + 4 \cT \left(1-\cT^2-\es^2\right) \Xtz \left(\Xtdm + \Xtdp\right) 
 - 2 \cT \left(1-\cT^2-4 \es^2\right) \Xtdp \Xtdm \bigg]  \\
&  + \dfrac{3}{16} \, \bd  \bigg[   
   2 \cT \left(1-\cT^2\right) \left(\Xtz\right)^2
 + \dfrac12 \left(1-\cT\right)^3 \left(\Xtdm\right)^2
 - \dfrac12 \left(1+\cT\right)^3 \left(\Xtdp\right)^2 \\&
 + \left(\left(1-\cT\right)^2 \left(1+2 \cT\right)-2 \left(1-\cT\right) \es^2\right) \Xtz \Xtdm
 - \left(\left(1+\cT\right)^2 \left(1-2 \cT\right)-2 \left(1+\cT\right) \es^2\right) \Xtz \Xtdp    \\&
 + \cT \left(1-\cT^2-4 \es^2\right) \Xtdp \Xtdm \bigg] \Bigg\} \ , 
\end{split}
\ee
\be
\llabel{211027t2}
\small
\begin{split}
T_2 & =
%{\sf T2c} = 
\At \sumk \Bigg\{  \dfrac{3}{32} \, \bz  \, \cT \bigg[
   3 \left(1-\cT^2\right) \left(\left(\Xtdm\right)^2 - \left(\Xtdp\right)^2\right) \\
& - 2 \left(1-3 \cT^2-6 \es^2\right) \Xtz \left(\Xtdm - \Xtdp\right) 
 \bigg] \\
&  + \dfrac{3}{16} \, \bu  \bigg[   
   4 \cT^2 \left(\Xtz\right)^2
 + \left(1-\cT\right)^2 \left(1+2 \cT\right) \left(\Xtdm\right)^2
 + \left(1+\cT\right)^2 \left(1-2 \cT\right) \left(\Xtdp\right)^2 \\&
 + 4 \es^2 \Xtz \left(\Xtdm + \Xtdp\right)
 + 4 \cT \left(1-\cT^2-2 \es^2\right) \Xtz \left(\Xtdm - \Xtdp\right) \\
& - 2 \left(1-\cT^2 -4 \es^2\right) \Xtdp \Xtdm  \bigg]  \\
&  + \dfrac{3}{32} \, \bd  \bigg[   
   4 \left(1-\cT^2\right) \left(\Xtz\right)^2
 + \left(1-\cT\right)^3 \left(\Xtdm\right)^2
 + \left(1+\cT\right)^3 \left(\Xtdp\right)^2 \\&
 + 4 \left(1-\es^2\right) \Xtz \left(\Xtdm + \Xtdp\right) 
 - 2 \cT \left(3-\cT^2-2 \es^2\right) \Xtz \left(\Xtdm - \Xtdp\right)  \\&
 + 2 \left(1-\cT^2-4 \es^2\right) \Xtdp \Xtdm
 \bigg] \Bigg\} \ , 
\end{split}
\ee
\be
\llabel{211027t3}
\small
\begin{split}
T_3 & = 
%\frac{\sf Txc[1]}{\sT} =
- \At \sumk  \Bigg\{  \dfrac{3}{16} \, \cT \, \az  \bigg[
   2 \left(1-3 \cT^2\right) \left(\Xtz\right)^2
 + \dfrac32 \left(1-\cT^2\right) \left(\left(\Xtdm\right)^2 + \left(\Xtdp\right)^2\right) \\&
 - 2 \left(2-3 \left(\cT^2+\es^2\right)\right) \Xtz \left(\Xtdm + \Xtdp\right)
 + 3 \left(1-\cT^2-4 \es^2\right) \Xtdp \Xtdm \bigg]  \\
&  - \dfrac{3}{16} \, \au  \bigg[   
   4 \cT \left(1-2 \cT^2\right) \left(\Xtz\right)^2
 - \left(1-\cT\right)^2 \left(1+2 \cT\right) \left(\Xtdm\right)^2 
 + \left(1+\cT\right)^2 \left(1-2 \cT\right) \left(\Xtdp\right)^2 \\&
 + 2 \left(\left(1-\cT\right) \left(1-\cT-4 \cT^2\right)-2 \left(1-2 \cT\right) \es^2\right) \Xtz \Xtdm \\&
 - 2 \left(\left(1+\cT\right) \left(1+\cT-4 \cT^2\right)-2 \left(1+2 \cT\right) \es^2\right) \Xtz \Xtdp 
 + 4 \cT \left(1-\cT^2-4 \es^2\right) \Xtdp \Xtdm \bigg]  \\
&  + \dfrac{3}{16} \, \ad  \bigg[   
   2 \cT \left(1-\cT^2\right) \left(\Xtz\right)^2
 + \dfrac12 \left(1-\cT\right)^3 \left(\Xtdm\right)^2 
 - \dfrac12 \left(1+\cT\right)^3 \left(\Xtdp\right)^2 \\&
 + \left(\left(1-\cT\right)^2 \left(1+2 \cT\right)-2 \left(1-\cT\right) \es^2\right) \Xtz \Xtdm 
 - \left(\left(1+\cT\right)^2 \left(1-2 \cT\right)-2 \left(1+\cT\right) \es^2\right) \Xtz \Xtdp  \\&
 + \cT \left(1-\cT^2-4 \es^2\right) \Xtdp \Xtdm \bigg] \Bigg\} \ , 
\end{split}
\ee
\be
\llabel{211027t4}
\small
\begin{split}
T_4 & =
%{\sf T4c} = 
- \At \sumk \Bigg\{ \dfrac{3}{4} \, \bz  \, \cT \es \, \Xtz \left(\Xtdm - \Xtdp\right)  \\
&  + \dfrac{3}{4} \, \es \, \bu  \bigg[   
   \left(1-\cT^2\right) \Xtz \left(\Xtdm + \Xtdp\right)  
 - 2 \left(1-\cT^2-2 \es^2\right) \Xtdp \Xtdm \bigg]  \\
&  + \dfrac{3}{8} \, \es \, \bd  \bigg[   
   \left(1-\cT\right)^2 \Xtz \Xtdm 
 + \left(1+\cT\right)^2 \Xtz \Xtdp 
 + 2 \left(1-\cT^2-2 \es^2\right) \Xtdp \Xtdm \bigg] \Bigg\} \ , 
\end{split}
\ee
\be
\llabel{211027t5}
\small
\begin{split}
T_5 & =
%{\sf T5c} = 
- \At \sumk \Bigg\{  \dfrac{3}{8} \, \es \, \az  \bigg[
   \left(1-3 \cT^2\right) \Xtz \left(\Xtdm + \Xtdp\right)  
 - 6 \left(1-\cT^2-2 \es^2\right) \Xtdp \Xtdm \bigg]  \\
&  - \dfrac{3}{2} \, \es \, \au  \bigg[   
   \cT \left(1-\cT\right) \Xtz \Xtdm 
 - \cT \left(1+\cT\right) \Xtz \Xtdp 
 - 2 \left(1-\cT^2-2 \es^2\right) \Xtdp \Xtdm \bigg]  \\
&  - \dfrac{3}{8} \, \es \, \ad  \bigg[   
   \left(1-\cT\right)^2 \Xtz \Xtdm 
 + \left(1+\cT\right)^2 \Xtz \Xtdp 
 + 2 \left(1-\cT^2-2 \es^2\right) \Xtdp \Xtdm
 \bigg] \Bigg\} \ ,
\end{split}
\ee
where
\be
\At =  \frac{\Gc \Ms^2 R^5}{a^6}
%=  \beta n a^2 n \frac{\Ms}{\ms} \left(\frac{R}{a}\right)^5
%=  \beta n a^2 \Ae
\ . \llabel{211029c}
\ee

We note that terms in $\es = \ue \cdot \vs$ only appear combined with the product of two different Hansen coefficients.
This product is proportional to $e^2$ (Table~\ref{tabHansen}) and therefore there is no problem if we are unable to accurately obtain $\ue$ from $\ve$ to compute $\es$ when $e \approx 0$. % (Eq.\,(\ref{211026a})).
We also note that the coefficients $T_1$, $T_2$, and $T_4$ solely depend on $b(\sigma)$, thus contributing to a secular evolution of the angular momenta, while the coefficients $T_3$ and $T_5$ solely depend on $a(\sigma)$, thus resulting only in a precession of the angular momentum vectors.

\subsection{Laplace vector}

The Laplace vector is responsible for the variations that occur in the eccentricity and argument of the pericentre (Eq.\,(\ref{210805c})).
As for the torque (Eq.\,(\ref{211017a})), we first need to express $\de$ in the reference frame $(\vp, \vq, \vs)$, such that we can write the expression of $\TI$ using Love numbers and the Hansen coefficients (Eq.\,(\ref{211001f})).
However, since the Laplace vector is attached to the orbital plane (Fig.~\ref{frames}), its expression becomes simpler if we project it on the frame $(\ue, \vx, \vk)$, which can be done by inverting expression (\ref{210804d}).
Moreover, {two of the projections in this new frame directly give the variation of the eccentricity and the argument of the pericentre}. %, which can be interesting to study some problems. 
Therefore, after averaging over the mean anomaly and rewriting the Hansen coefficients in the format $X_k^{-3,m}$ (appendix~\ref{hcr}), we get
\be
\big\langle \de \big\rangle_M = \dot e \, \ue + \dot \ek \, \vk + e \dot \vpi \, \vk \times \ue  
\ , \llabel{211026b}
\ee
with
\be
\llabel{211026z1}
\small
\begin{split}
\dot e & = -\Ae \, \frac{\sq}{e} \sumk \Bigg\{  \dfrac38 \, \es \eks \, \az  \bigg[
  \left(1-3 \cT^2\right) \Xtz \left(\Xtdm + \Xtdp\right) %\\& 
  - 6\left(1-\cT^2-2\es^2\right) \Xtdp \Xtdm \bigg]  \\
& - \dfrac32 \, \es \eks \, \au  \bigg[
   \cT \left(1-\cT\right) \Xtz \Xtdm
 - \cT \left(1+\cT\right) \Xtz \Xtdp 
 - 2 \left(1-\cT^2-2\es^2\right) \Xtdp \Xtdm \bigg] \\
 &  - \dfrac38 \, \es \eks \, \ad  \bigg[
    \left(1-\cT\right)^2 \Xtz \Xtdm
 + \left(1+\cT\right)^2 \Xtz \Xtdp 
 + 2 \left(1-\cT^2-2\es^2\right) \Xtdp \Xtdm \bigg]  \\
 &  - \dfrac{1}{32} \, \bz  \bigg[ 
   2 \left(1-3\cT^2\right)^2 \left(\Xtz\right)^2 k \sq %\\&
 + 9 \left(1-\cT^2\right)^2 \left(\Xtdm\right)^2 \left(1+ \frac{k}{2} \sq\right)  \\&
 - 9 \left(1-\cT^2\right)^2 \left(\Xtdp\right)^2 \left(1- \frac{k}{2} \sq\right)  %\\ &
 - 6 \left(1-3\cT^2\right) \left(1-\cT^2-2\es^2\right) \Xtz \Xtdm \left(1+k \sq\right)  \\ &
 + 6 \left(1-3\cT^2\right) \left(1-\cT^2-2\es^2\right) \Xtz \Xtdp \left(1-k \sq\right)  %\\ &
 + 9 \left(\left(1-\cT^2\right)^2-8 \left(1-\cT^2-\es^2\right)\es^2\right) \Xtdp \Xtdm k \sq \bigg]  \\
 &  - \dfrac{3}{16} \, \bu  \bigg[   
    4 \left(1-\cT^2\right) \cT^2 \left(\Xtz\right)^2 k \sq  %\\&
 + \left(1-\cT^2\right) \left(1-\cT\right)^2 \left(\Xtdm\right)^2 \left(2+k \sq\right)  \\&
 - \left(1-\cT^2\right) \left(1+\cT\right)^2 \left(\Xtdp\right)^2 \left(2-k \sq\right)  %\\&
 + 4\cT \left(1-\cT\right) \left(1-\cT^2-2 \es^2\right) \Xtz \Xtdm \left(1+k \sq\right)  \\&
 + 4\cT \left(1+\cT\right) \left(1-\cT^2-2 \es^2\right) \Xtz \Xtdp \left(1-k \sq\right)  \\&
 - 2 \left(\left(1-\cT^2\right)^2-8 \left(1-\cT^2-\es^2\right) \es^2\right) \Xtdp \Xtdm k \sq \bigg]  \\
  &  - \dfrac{3}{64} \, \bd  \bigg[   
 4 \left(1-\cT^2\right)^2 \left(\Xtz\right)^2 k \sq   \\&
 + \left(1-\cT\right)^4 \left(\Xtdm\right)^2 \left(2+k \sq\right)   
 - \left(1+\cT\right)^4 \left(\Xtdp\right)^2 \left(2-k \sq\right)   \\&
 + 4 \left(1-\cT\right)^2 \left(1-\cT^2-2 \es^2\right) \Xtz \Xtdm \left(1+k \sq\right)   \\&
 - 4 \left(1+\cT\right)^2 \left(1-\cT^2-2 \es^2\right) \Xtz \Xtdp \left(1-k \sq\right)   \\&
 + 2 \left(\left(1-\cT^2\right)^2-8 \left(1-\cT^2-\es^2\right) \es^2\right) \Xtdp \Xtdm k \sq \bigg] \Bigg\} \ ,
\end{split}
\ee

\be
\llabel{211026z2}
\small
\begin{split}
\dot \ek & = - \frac{ \Ae e}{\sq} \sumk  \Bigg\{  \dfrac{3}{32} \, \eks \, \az \bigg[
   4 \cT \left(1-3 \cT^2\right) \left(\Xtz\right)^2
 + 3 \cT \left(1-\cT^2\right) \left(\left(\Xtdm\right)^2 + \left(\Xtdp\right)^2\right) \\&
 - 4 \cT \left(2-3 \left(\cT^2+\es^2\right)\right) \Xtz \left(\Xtdm + \Xtdp\right)
 + 6 \cT \left(1-\cT^2-4 \es^2\right) \Xtdp \Xtdm \bigg]  \\
 &  - \dfrac{3}{16} \, \eks \, \au \bigg[
   4 \cT \left(1-2 \cT^2\right) \left(\Xtz\right)^2
 - \left(1-\cT\right)^2 \left(1+2 \cT\right) \left(\Xtdm\right)^2
 + \left(1+\cT\right)^2 \left(1-2 \cT\right) \left(\Xtdp\right)^2  \\&
 + 2 \left(\left(1-\cT\right) \left(1-\cT-4 \cT^2\right)-2 \left(1-2 \cT\right) \es^2\right) \Xtz \Xtdm \\&
 - 2 \left(\left(1+\cT\right) \left(1+\cT-4 \cT^2\right)-2 \left(1+2 \cT\right) \es^2\right) \Xtz \Xtdp 
 + 4 \cT \left(1-\cT^2-4 \es^2\right) \Xtdp \Xtdm \bigg]  \\
  &  + \dfrac{3}{32} \, \eks \, \ad \bigg[
     4 \cT \left(1-\cT^2\right) \left(\Xtz\right)^2
 + \left(1-\cT\right)^3 \left(\Xtdm\right)^2
 - \left(1+\cT\right)^3 \left(\Xtdp\right)^2 \\&
 + 2 \left(\left(1-\cT\right)^2 \left(1+2 \cT\right)-2 \left(1-\cT\right) \es^2\right) \Xtz \Xtdm
 - 2 \left(\left(1+\cT\right)^2 \left(1-2 \cT\right)-2 \left(1+\cT\right) \es^2\right) \Xtz \Xtdp \\&
 + 2 \cT \left(1-\cT^2-4 \es^2\right) \Xtdp \Xtdm \bigg]  \\
  &  + \dfrac{3}{32} \, \es \, \bz \bigg[
     3 \cT \left(1-\cT^2\right) \left(\left(\Xtdm\right)^2 - \left(\Xtdp\right)^2\right)
 - 2 \cT \left(5- 3 \left(\cT^2+2 \es^2\right)\right) \Xtz \left(\Xtdm - \Xtdp\right)  \bigg]  \\
  &  + \dfrac{3}{16} \, \es \, \bu \bigg[
   4 \cT^2 \left(\Xtz\right)^2
 + \left(1-\cT\right)^2 \left(1+2 \cT\right) \left(\Xtdm\right)^2
 + \left(1+\cT\right)^2 \left(1-2 \cT\right) \left(\Xtdp\right)^2 \\&
 - 4 \left(\left(1-\cT\right) \left(1-\cT^2\right)-\left(1-2 \cT\right) \es^2\right) \Xtz \Xtdm
 - 4 \left(\left(1+\cT\right) \left(1-\cT^2\right)-\left(1+2 \cT\right) \es^2\right) \Xtz \Xtdp \\&
 + 2 \left(3-3 \cT^2-4 \es^2\right) \Xtdp \Xtdm  \bigg]  \\
  &  + \dfrac{3}{32} \, \es \, \bd \bigg[
   4 \left(1-\cT^2\right) \left(\Xtz\right)^2
 + \left(1-\cT\right)^3 \left(\Xtdm\right)^2
 + \left(1+\cT\right)^3 \left(\Xtdp\right)^2 \\&
 + 2 \left(\cT \left(1-\cT\right)^2-2 \left(1-\cT\right) \es^2\right) \Xtz \Xtdm
 - 2 \left(\cT \left(1+\cT\right)^2+2 \left(1+\cT\right) \es^2\right) \Xtz \Xtdp \\&
 - 2 \left(3-3\cT^2-4 \es^2\right) \Xtdp \Xtdm \bigg] \Bigg\}  \ ,
\end{split}
\ee

\be
\llabel{211026z3}
\small
\begin{split}
e \dot \vpi_a & = - \frac{\Ae }{e \sq}  \sumk \Bigg\{  \dfrac{3}{32} \, \az \bigg[
   2 \left(\cT^2-\es^2-3 \cT^2 \left(\cT^2-\es^2\right)\right) e^2 (\Xtz)^2 \\&
 + \dfrac32 \left(\left(1-\cT^2\right)^2 \left(2+k \left(1-e^2\right)^{3/2}\right)-\left(\cT^2 \left(1-\cT^2\right)+\left(3-5 \cT^2-2 \es^2\right) \es^2\right) e^2\right) \left(\Xtdm\right)^2 \\&
 + \dfrac32 \left(\left(1-\cT^2\right)^2 \left(2-k \left(1-e^2\right)^{3/2}\right)-\left(\cT^2 \left(1-\cT^2\right)+\left(3-5 \cT^2-2 \es^2\right) \es^2\right) e^2\right) \left(\Xtdp\right)^2 \\&
 - 2 \left(\left(1-\cT^2 \left(4-3 \cT^2\right)-2 \es^2 \left(1-3 \cT^2\right)\right)+\left(\cT^2-\es^2\right) \left(1-3 \es^2\right) e^2\right) \Xtz \left(\Xtdm + \Xtdp\right) \\&
 - \left(1-\cT^2 \left(4-3 \cT^2\right)-2 \es^2 \left(1-3 \cT^2\right)+4 \es^2 \left(1-\cT^2-\es^2\right)\right) \Xtz \left(\Xtdm - \Xtdp\right) k \left(1-e^2\right)^{3/2}   \\&
 + 3 \left(2 \left(1-\cT^2\right)^2-16 \es^2 \left(1-\cT^2-\es^2\right)+\left(\es^2 \left(5-6 \es^2\right)-\cT^2 \left(1-\cT^2+3 \es^2\right)\right) e^2\right) \Xtdp \Xtdm \\&
 - 2 \left(1-\es^2-\cT^2 \left(5-6 \cT^2-3 \es^2\right)\right) \Xtz \left(\Xtum + \Xtup\right) e \\&
 + 3 \left(\left(1-3 \es^2+2 \es^4\right)-\cT^2 \left(3-2 \cT^2-5 \es^2\right)\right) \left(\Xtdm + \Xtdp\right) \left(\Xtum + \Xtup\right) e \bigg]   \\
  &  - \dfrac{1}{8} \, \au \bigg[ 
   2 \cT \left(3 \cT \left(1-\cT^2+\es^2\right) e^2+\left(1-\cT^2-2 \es^2\right) k \left(1-e^2\right)^{3/2}\right) (\Xtz)^2  \\&
 - \dfrac32 \left(\left(1-\cT\right)^2 \left(1-\cT^2\right) \left(2+k \left(1-e^2\right)^{3/2}\right)-\left(1-2 \es^2 \left(1-\es^2\right)-3 \cT \es^2-\cT^2 \left(2-5 \es^2-\cT^2\right)\right) e^2\right) \left(\Xtdm\right)^2 \\&
 - \dfrac32 \left(\left(1+\cT\right)^2 \left(1-\cT^2\right) \left(2-k \left(1-e^2\right)^{3/2}\right)-\left(1-2 \es^2 \left(1-\es^2\right)+3 \cT \es^2-\cT^2 \left(2-5 \es^2-\cT^2\right)\right) e^2\right) \left(\Xtdp\right)^2 \\&
 + 2 \left(3 \cT^2 \left(1-\cT^2-2 \es^2\right)+3 \es^2 \left(1+\cT^2-\es^2\right) e^2-\cT \left(2-2 \cT^2-3 \es^2\right) k \left(1-e^2\right)^{3/2}\right) \Xtz \left(\Xtdm + \Xtdp\right) \\&
 - \Big(6 \cT \left(1-\cT^2-2 \es^2\right)-3 \cT \left(2-2 \cT^2-5 \es^2\right) e^2
  \\& \quad\quad\quad\quad\quad\quad\quad\quad
 -\left(1-4 \es^2 \left(1-\es^2\right)+\cT^2 \left(2-3 \cT^2-2 \es^2\right)\right) k \left(1-e^2\right)^{3/2}\Big) \Xtz \left(\Xtdm - \Xtdp\right) \\&
 + 3 \left(2 \left(1-\cT^2\right)^2-16 \es^2 \left(1-\cT^2-\es^2\right)-\left(1-6 \es^2 \left(1-\es^2\right)+\cT^2 \left(3 \es^2-\cT^2\right)\right) e^2\right) \Xtdp \Xtdm \\&
 + 6 \cT^2 \left(2-2 \cT^2-\es^2\right) \Xtz \left(\Xtum + \Xtup\right) e \\&
 - 3 \left(2 \es^2 \left(1-\es^2\right)-\cT \left(2-3 \es^2\right)+\cT^2 \left(2+2 \cT-2 \cT^2-5 \es^2\right)\right) \Xtdm \left(\Xtum + \Xtup\right)  e \\&
 - 3 \left(2 \es^2 \left(1-\es^2\right)+\cT \left(2-3 \es^2\right)+\cT^2 \left(2-2 \cT-2 \cT^2-5 \es^2\right)\right) \Xtdp \left(\Xtum + \Xtup\right)  e \bigg] \\
  & + \dfrac{1}{32} \, \ad \bigg[ 
   2 \left(2 \cT \left(1-\cT^2-2 \es^2\right) k \left(1-e^2\right)^{3/2}-3 \left(2-\es^2-\cT^2 \left(3-\cT^2+\es^2\right)\right) e^2\right) (\Xtz)^2 \\&
 + \dfrac32 \left(\left(1-\cT\right)^4 \left(2+k \left(1-e^2\right)^{3/2}\right)-\left(2-\es^2 \left(1+2 \es^2\right)-6 \cT \left(1-\es^2\right)+\cT^2 \left(5-5 \es^2-\cT^2\right)\right) e^2\right) \left(\Xtdm\right)^2 \\&
 + \dfrac32 \left(\left(1+\cT\right)^4 \left(2-k \left(1-e^2\right)^{3/2}\right)-\left(2-\es^2 \left(1+2 \es^2\right)+6 \cT \left(1-\es^2\right)+\cT^2 \left(5-5 \es^2-\cT^2\right)\right) e^2\right) \left(\Xtdp\right)^2 \\&
 + 2 \Big(3 \left(1-2 \es^2-\cT^2 \left(\cT^2+2 \es^2\right)\right)-3 \left(2-\es^2 \left(3-\es^2\right)-\cT^2 \left(1+\es^2\right)\right) e^2 
  \\& \quad\quad\quad\quad\quad\quad\quad\quad
 -2 \cT \left(1-3 \es^2-2 \cT^2\right) k \left(1-e^2\right)^{3/2}\Big) \Xtz \left(\Xtdm + \Xtdp\right) \\&
 - \Big(12 \cT \left(1-\cT^2-2 \es^2\right)-6 \cT \left(3-2 \cT^2-5 \es^2\right) e^2
  \\& \quad\quad\quad\quad\quad\quad\quad\quad
 -\left(3-2 \es^2 \left(5-2 \es^2\right)-\cT^2 \left(4+2 \es^2+3 \cT^2\right)\right) k \left(1-e^2\right)^{3/2}\Big) \Xtz \left(\Xtdm - \Xtdp\right) \\&
 + 3 \left(2 \left(1-\cT^2\right)^2-16 \es^2 \left(1-\cT^2-\es^2\right)-\left(2-9 \es^2+6 \es^4-\cT^2 \left(3+\cT^2-3 \es^2\right)\right) e^2\right) \Xtdp \Xtdm  \\&
 - 6 \left(1+\es^2-\cT^2 \left(3-2 \cT^2-\es^2\right)\right) \Xtz \left(\Xtum + \Xtup\right) e \\&
 - 3 \left(1-\es^2 \left(1+2 \es^2\right)-2 \cT \left(1-3 \es^2\right)-\cT^2 \left(1-4 \cT+2 \cT^2+5 \es^2\right)\right) \Xtdm \left(\Xtum + \Xtup\right)  e \\&
 - 3 \left(1-\es^2 \left(1+2 \es^2\right)+2 \cT \left(1-3 \es^2\right)-\cT^2 \left(1+4 \cT+2 \cT^2+5 \es^2\right)\right) \Xtdp \left(\Xtum + \Xtup\right)  e  \bigg] \Bigg\} \ ,
\end{split}
\ee

\be
\small
\begin{split}
e \dot \vpi_b & = \frac{\Ae}{e \sq}  \sumk \Bigg\{ \dfrac{1}{16} \, \es \eks \, \bz \bigg[ 
   2 \left(1-3 \cT^2\right) (\Xtz)^2 k \left(1-e^2\right)^{3/2} \\&
 - \dfrac92 \left(1-2 \cT^2-\es^2\right) e^2 \left(\left(\Xtdm\right)^2 - \left(\Xtdp\right)^2\right) \\&
 - 3 \left(2 \left(1-3 \cT^2\right)-\left(1-3 \es^2\right) e^2\right) \Xtz \left(\Xtdm - \Xtdp\right) \\&
 - 6 \left(1-2 \cT^2-\es^2\right) \Xtz \left(\Xtdm + \Xtdp\right) k \left(1-e^2\right)^{3/2} \\&
 + 18 \left(1-\cT^2-2 \es^2\right) \Xtdp \Xtdm k \left(1-e^2\right)^{3/2} \\&
 - 9 \left(1-2 \cT^2-\es^2\right) \left(\Xtum + \Xtup\right) \left(\Xtdm - \Xtdp\right) e \bigg] \\
  & - \dfrac{1}{8} \, \es \eks \, \bu \bigg[ 
   2 \cT \left(3 e^2-2 \cT \, k \left(1-e^2\right)^{3/2}\right) (\Xtz)^2 \\&
 - \dfrac32 \left(1+3 \cT-4 \cT^2-2 \es^2\right) e^2 \left(\Xtdm\right)^2 
 + \dfrac32 \left(1-3 \cT-4 \cT^2-2 \es^2\right) e^2 \left(\Xtdp\right)^2 \\&
 - 2 \left(6 \cT+3 \cT e^2+\left(1-4 \cT^2-2 \es^2\right) k \left(1-e^2\right)^{3/2}\right) \Xtz \left(\Xtdm + \Xtdp\right) \\&
 + 3 \left(4 \cT^2+\left(1-2 \es^2\right) e^2-2 \cT \, k \left(1-e^2\right)^{3/2}\right) \Xtz \left(\Xtdm - \Xtdp\right) \\&
 + 3 \left(4 \left(1-\cT^2-2 \es^2\right) k \left(1-e^2\right)^{3/2}-3 \cT e^2\right) \Xtdp \Xtdm 
 - 6 \cT \, \Xtz \left(\Xtum + \Xtup\right) e \\&
 - 3 \left(1+3 \cT-4 \cT^2-2 \es^2\right) \Xtdm \left(\Xtum + \Xtup\right)  e \\&
 + 3 \left(1-3 \cT-4 \cT^2-2 \es^2\right) \Xtdp \left(\Xtum + \Xtup\right)  e \bigg] \\
  & - \dfrac{1}{16} \, \es \eks \, \bd \bigg[ 
   2 \left(\left(1+\cT^2\right) k \left(1-e^2\right)^{3/2}-3 \cT e^2\right) (\Xtz)^2 \\&
 - \dfrac32 \left(1-3 \cT+2 \cT^2+\es^2\right) e^2 \left(\Xtdm\right)^2 
 + \dfrac32 \left(1+3 \cT+2 \cT^2+\es^2\right) e^2 \left(\Xtdp\right)^2 \\&
 + 2 \left(6 \cT+3 \cT e^2-\left(1+2 \cT^2+\es^2\right) k \left(1-e^2\right)^{3/2}\right) \Xtz \left(\Xtdm + \Xtdp\right) \\&
 - 3 \left(2 \left(1+\cT^2\right)+\left(1-\es^2\right) e^2-2 \cT \, k \left(1-e^2\right)^{3/2}\right) \Xtz \left(\Xtdm - \Xtdp\right) \\&
 - 3 \left(2 \left(1-\cT^2-2 \es^2\right) k \left(1-e^2\right)^{3/2}-3 \cT \, e^2\right) \Xtdp \Xtdm 
 + 6 \cT \, \Xtz \left(\Xtum + \Xtup\right) e \\&
 - 3 \left(1-3 \cT+2 \cT^2+\es^2\right) \Xtdm \left(\Xtum + \Xtup\right)  e \\&
 + 3 \left(1+3 \cT+2 \cT^2+\es^2\right) \Xtdp \left(\Xtum + \Xtup\right)  e \bigg] \Bigg\} \ ,
\end{split}
\ee
where
\be
\dot \vpi = \dot \vpi_a + \dot \vpi_b
 \ , \llabel{211026y}
\ee
and
\be
\Ae = n \left( \frac{\Ms}{\ms} \right) \left( \frac{R}{a} \right)^5  
= \frac{\At}{\beta n a^2} \ .
\ee

We note that for the $\dot e$ and $e \dot \vpi$ projections of the Laplace vector, all terms in the series appear combined with one of the following combinations of the Hansen coefficients: $X_k^{-3,m} X_k^{-3,m'}$ with $m \ne m'$, $e X_k^{-3,\pm 1} X_k^{-3,m}$, $e^2 (X_k^{-3,m})^2$, $k (X_k^{-3,0})^2$, or $(2\mp k) (X_k^{-3,\pm 2})^2$.
As a result, all these terms are proportional to $e^2$ for all values of $k$ (Table~\ref{tabHansen}), which suppresses the potential singularity at $e=0$ introduced by the factor $e^{-1}$ in their amplitudes.
It also ensures that there is no problem with the terms in $\es$ and $\eks$ if we are unable to accurately obtain $\ue$ from $\ve$ when $e \approx 0$ (Eq.\,(\ref{211026a})).

\subsection{Orbital and spin evolution}
\llabel{oase1}

The set of equations (\ref{211017d}) and (\ref{211026b}) allow us to track the evolution of the averaged system using the angular momentum and the Laplace vectors (Sect.~\ref{pmp}).
For a more intuitive description of the orbital and spin evolution, we can relate these vectors with the rotation and orbital elliptic elements.
The eccentricity can be directly obtained from the Laplace vector,
\be
e = \sqrt{\ve \cdot \ve} 
%\ , \quad \mathrm{and}  \quad
%\cos \vpi = \ue \cdot \vp = \frac{\ve \cdot \vp}{e} = 
\ , \llabel{211027a}
\ee
the semi-major axis from the orbital angular momentum (Eq.\,(\ref{210804a})),
\be
a %= \frac{(\vG \cdot \vk)^2}{\mu \beta^2 (1-e^2)}
= \frac{\vG \cdot \vG}{\beta^2 \mu (1-e^2)}
\ , \llabel{211027b}
\ee
and the rotation rate from the rotational angular momentum (Eq.\,(\ref{151019b})),
\be
\om %= \frac{\vL \cdot \vs}{C} 
= \frac{\sqrt{\vL \cdot \vL}}{C}
\ . \llabel{211027c}
\ee
The argument of the pericentre is given by expression (\ref{210915a}), while the angle between the orbital and equatorial planes, can be obtained from both angular momentum vectors as 
\be
\cos \theta = \vk \cdot \vs = \frac{\vG \cdot \vL}{\sqrt{(\vG \cdot \vG) (\vL \cdot \vL)}} 
\ . \llabel{211027z}
\ee

For a better comparison with previous studies, we can also obtain the evolution of all these quantities. The eccentricity and the argument of the pericentre evolution are already given by two projections of the Laplace vector, namely by $\dot e$ (Eq.\,(\ref{211026z1})) and $\dot \vpi$ (Eq.\,(\ref{211026y})), respectively:
\be
\dot e = \de \cdot \ue \ , \quad \mathrm{and} \quad \dot \vpi = \de \cdot (\vk \times \ue) / e
\ . \llabel{211027d}
\ee
The semi-major axis evolution is given from expressions (\ref{150626a}), (\ref{211027b}) and (\ref{211027d}),
\be
\dot a = \frac{2 \, \vG \cdot \vT}{\beta^2 \mu (1-e^2)} + \frac{2 a \, \ve \cdot \de}{(1-e^2)}
= \frac{2 (T_1 + T_2 \, \cT - T_5 \, \eks)}{\beta n a \sqrt{1-e^2}} + \frac{2 a e \dot e}{(1-e^2)} 
\ , \llabel{211027e}
\ee
or, making use of expressions (\ref{211027t1}), (\ref{211027t2}), (\ref{211027t5}), and (\ref{211026z1}), we have
\be
\llabel{211028c}
\small
\begin{split}
\frac{\dot a}{a} & =
%\frac{\sf TMabc[4]}{n} = & 
 \Ae \sumk \Bigg\{ \dfrac{1}{16} \, k \,\bz  \bigg[
   2 \left(1-3 \cT^2\right)^2 \left(\Xtz\right)^2
 + \dfrac92 \left(1-\cT^2\right)^2 \left(\left(\Xtdm\right)^2 + \left(\Xtdp\right)^2\right) \\&
 - 6 \left(1-3 \cT^2\right) \left(1-\cT^2-2 \es^2\right) \Xtz \left(\Xtdm + \Xtdp\right)
 + 9 \left(\left(1-\cT^2\right)^2-8 \left(1-\cT^2-\es^2\right) \es^2\right) \Xtdp \Xtdm
 \bigg]  \\
&  + \dfrac{3}{8} \, k \, \bu  \bigg[   
   4 \left(1-\cT^2\right) \cT^2 \left(\Xtz\right)^2
 + \left(1-\cT^2\right) \left(1-\cT\right)^2 \left(\Xtdm\right)^2
 + \left(1-\cT^2\right) \left(1+\cT\right)^2 \left(\Xtdp\right)^2 \\&
 + 4 \cT \left(1-\cT\right) \left(1-\cT^2-2 \es^2\right) \Xtz \Xtdm
 - 4 \cT \left(1+\cT\right) \left(1-\cT^2-2 \es^2\right) \Xtz \Xtdp \\&
 - 2 \left(\left(1-\cT^2\right)^2-8 \left(1-\cT^2-\es^2\right) \es^2\right) \Xtdp \Xtdm
 \bigg]  \\
&  + \dfrac{3}{32} \, k \, \bd  \bigg[   
   4 \left(1-\cT^2\right)^2 \left(\Xtz\right)^2
 + \left(1-\cT\right)^4 \left(\Xtdm\right)^2
 + \left(1+\cT\right)^4 \left(\Xtdp\right)^2 \\&
 + 4 \left(1-\cT\right)^2 \left(1-\cT^2-2 \es^2\right) \Xtz \Xtdm
 + 4 \left(1+\cT\right)^2 \left(1-\cT^2-2 \es^2\right) \Xtz \Xtdp \\&
 + 2 \left(\left(1-\cT^2\right)^2-8 \left(1-\cT^2-\es^2\right) \es^2\right) \Xtdp \Xtdm
 \bigg] \Bigg\} \ .
\end{split}
\ee

For the rotation rate evolution, we have from expressions (\ref{210805b}) and (\ref{211027c}),
\be
\dot \om = - \frac{\vT \cdot \vs}{C} 
%= - \frac{T_s}{C} 
= - \frac{T_1 \, \cT + T_2 + T_4 \, \es}{C} 
\ , \llabel{211028a}
\ee
that is, using expressions (\ref{211027t1}), (\ref{211027t2}), and (\ref{211027t4}), 
\be
\llabel{211028b}
\small
\begin{split}
\dot \om & =
%- \frac{1}{C} \frac{\sf TMabc[3]} = & 
- \frac{\At}{C} \sumk \Bigg\{ \dfrac{3}{16} \, \bu  \bigg[  
   \left(1-\cT\right)^2 \left(1-\cT^2\right) \left(\Xtdm\right)^2 
 + \left(1+\cT\right)^2 \left(1-\cT^2\right) \left(\Xtdp\right)^2 \\&
 - 2 \left(\left(1-\cT^2\right)^2-8 \left(1-\cT^2-\es^2\right) \es^2\right) \Xtdp \Xtdm
 + 4 \cT^2 \left(1-\cT^2\right) \left(\Xtz\right)^2 \\&
 + 4 \cT \left(1-\cT\right) \left(1-\cT^2-2 \es^2\right) \Xtz \Xtdm
 - 4 \cT \left(1+\cT\right) \left(1-\cT^2-2 \es^2\right) \Xtz \Xtdp 
 \bigg]  \\
&  + \dfrac{3}{32}  \, \bd  \bigg[  
   \left(1-\cT\right)^4 \left(\Xtdm\right)^2
 + \left(1+\cT\right)^4 \left(\Xtdp\right)^2 \\&
 + 2 \left(\left(1-\cT^2\right)^2-8 \left(1-\cT^2-\es^2\right) \es^2\right) \Xtdp \Xtdm
 + 4 \left(1-\cT^2\right)^2 \left(\Xtz\right)^2 \\&
 + 4 \left(1-\cT^2-2 \es^2\right) \left(1-\cT\right)^2 \Xtz \Xtdm
 + 4 \left(1-\cT^2-2 \es^2\right) \left(1+\cT\right)^2 \Xtz \Xtdp
  \bigg] \Bigg\} \ .
\end{split}
\ee
The obliquity (or inclination) evolution is given from expressions (\ref{150626a}), (\ref{210805b}) and (\ref{211027z}),
\be
\llabel{211028z}
\begin{split}
\dot \theta & = \frac{\vT \cdot \vk - \vT \cdot \vs \, \cos \theta}{\sin \theta \sqrt{\vL \cdot \vL}} - \frac{\vT \cdot \vs - \vT \cdot \vk \, \cos \theta}{\sin \theta \sqrt{\vG \cdot \vG}}  \\ &
% = \frac{(T_1 + T_2 \cT - T_5 \eks) - (T_1 \cT + T_2 + T_4 \es) \cT}{\sin \theta \sqrt{\vL \cdot \vL}} - \frac{(T_1 \cT + T_2 + T_4 \es) - (T_1 + T_2 \cT - T_5 \eks) \cT}{\sin \theta \sqrt{\vG \cdot \vG}} \\ &
 = \frac{T_1 \sin^2 \theta -  T_4 \, \cT \es - T_5 \, \eks}{C \om \sin \theta } - \frac{T_2 \sin^2 \theta + T_4 \, \es + T_5 \, \cT \eks}{\beta \sqrt{\mu a (1-e^2)} \sin \theta} \ .
\end{split}
\ee
We observe that there are two distinct contributions to the evolution of this angle.
When $| \vL | \ll | \vG |$, the evolution is dominated by the first term, and $\dot \theta$ is identified as a variation of the obliquity.
This is, for instance, the case of a planet around a star.
On the other hand, when $| \vL | \gg | \vG |$, the evolution is dominated by the second term, and $\dot \theta$ is recognised as a variation in the orbital inclination. 
This is, for instance, the case of a small satellite close to its planet.
Previous studies often neglect one of these two contributions, and therefore, the evolution of this angle is incomplete.
%A complete treatment of the evolution of this angle in terms of elliptical elements is provided by \citet{Boue_Efroimsky_2019}, which is equivalent to our expression (\ref{211028z}).

Finally, we can also obtain the evolution of the precession angles, that is, the angular velocity of the longitude of the node, $ \dot \Omega$, and the precession speed of the spin axis, $\dot \psi$.
The line of nodes is aligned with the vector $\vp$ (Fig.~\ref{frames}) and thus
%\be
%T_p = T_3 \sin \theta + T_4 \, \ue \cdot \vp - T_5 \, \ue \cdot \vq 
%= \frac{T_3 \sin^2 \theta - T_4 \eks  + T_5 \cT \es}{\sin \theta}
%\ee
\be
\dot \Omega = \frac{ \dot \vG}{|\vG|} \cdot \vp = \frac{\vT \cdot \vp}{|\vG|} 
=  \frac{T_3 \sin^2 \theta - T_4 \, \eks + T_5 \, \cT \es}{\beta \sqrt{\mu a (1-e^2)} \sin \theta}
\ , \llabel{211029a}
\ee
\be
\dot \psi = \frac{ \dot \vL}{|\vL|} \cdot \vp = - \frac{\vT \cdot \vp}{|\vL|} 
= - \frac{T_3 \sin^2 \theta - T_4 \, \eks + T_5 \, \cT \es}{C \om \sin \theta}
\ . \llabel{211029b}
\ee

\subsection{Energy dissipation}

The total energy released inside the body due to tides is given by
\be
\dot E = - (\dot E_{\rm orb} + \dot E_{\rm rot} )
\ , \llabel{211029d}
\ee
where
\be 
E_{\rm orb} = - \frac{\beta \mu}{2 a}
\quad \mathrm{and} \quad
E_{rot} = \frac{\vw \cdot \vL}{2}
\llabel{211029e}
\ee
are the orbital energy, and the rotational energy, respectively.
Then
\be
\dot E_{\rm orb} = \frac{\beta \mu}{2 a^2} \dot a 
\quad \mathrm{and} \quad
\dot E_{\rm rot} = C \om \dot \om
\ , \llabel{211029f}
\ee
where $\dot a$ and $\dot \om$ are given by expressions (\ref{211028c}) and (\ref{211028b}), respectively.

\section{Average over the argument of the pericentre}

\llabel{doubav}

Although we usually have $\dot \vpi \ll n$, the argument of the pericentre can often also be considered a fast varying angle when compared to the secular tidal evolution of the remaining spin and orbital elements. % $\dot \om$, $\dot \theta$, $\dot a$ and $\dot e$.
This is particularly true when additional sources of precession are taken into account in the problem, such as the rotational deformation, general relativity corrections and additional perturbing bodies in the system.
As a result, we can perform a second average of the equations of motion, this time over the angle $\vpi$.
Adopting again as example the last term in the $\vs$ component of the tidal torque, we have (Eq.\,(\ref{211006b}))
\be
\left\langle - \frac{3 \Gc \Ms}{r^3} I_{23} \xii \xk \right\rangle_{M,\vpi} 
% = \frac{3\,  \At }{16} \sin^2 \theta \sumk   
\!\!\! =  \frac{3 \Gc \Ms^2 R^5}{16 a^6}  \sin^2 \theta \sumk   
\bu \bigg[ (1-x) \left(X_k^{-3,-2}\right)^2 +  (1+x) \left(X_k^{-3,2}\right)^2  \bigg] 
\ , \llabel{211103f}
\ee
which considerably simplifies the expression of the equations of motion.
In this particular case, we note that there is no longer the contribution from $\au$.

Another simplification is that we do not need to follow the evolution of the Laplace vector anymore.
Indeed, the pericentre is no longer defined when we average over $\vpi$, and the only projection of interest is the one giving the evolution of the eccentricity, $\ue$.
Therefore, the equations of motion in this simplified case can be given by the torque together with $\dot e$.
Alternatively, we prefer to use the evolution of the orbital energy, since it can be obtained directly from the potential energy (Eq.\,(\ref{191014a})) and thus provides a simpler expression (from which we can later easily derive $\dot e$).

\subsection{Tidal torque}

When we average the tidal torque (Eq.\,(\ref{211017d})) over $\vpi$, the projections that depend on the Laplace vector average to zero and thus

\be
\big\langle \vT \big\rangle_{M,\vpi}   = \overline T_1 \, \vk + \overline T_2 \, \vs + \overline T_3 \, \vk \times \vs 
\ , \llabel{211110a}
\ee
with
\be
\llabel{211110t1}
\small
\begin{split}
\overline T_1 & = % \At  \frac{\sf TMwc[2]}{\sT} = 
  -  \At  \sumk \Bigg\{ \dfrac{9}{32} \, \bz  \bigg[
\left(1-\cT^2\right) \left(\left(\Xtdm\right)^2 - \left(\Xtdp\right)^2\right) \bigg]  \\
&  + \dfrac{3}{16} \, \bu  \bigg[   
   4 \cT^3 \left(\Xtz\right)^2
 + \left(1-\cT\right)^2 \left(2+\cT\right) \left(\Xtdm\right)^2
 - \left(1+\cT\right)^2 \left(2-\cT\right) \left(\Xtdp\right)^2
 \bigg]  \\
&  + \dfrac{3}{32} \, \bd  \bigg[   
   4 \cT \left(1-\cT^2\right) \left(\Xtz\right)^2
 + \left(1-\cT\right)^3 \left(\Xtdm\right)^2
 - \left(1+\cT\right)^3 \left(\Xtdp\right)^2
 \bigg] \Bigg\} \ ,
\end{split}
\ee
\be
\llabel{211110t2}
\small
\begin{split}
\overline T_2 & = % \At {\sf TMwc[5]} = 
 \At \sumk \Bigg\{ \dfrac{9}{32} \, \bz \bigg[
\cT \left(1-\cT^2\right) \left(\left(\Xtdm\right)^2 - \left(\Xtdp\right)^2\right) 
\bigg] \\
&  + \dfrac{3}{16} \, \bu  \bigg[   
   4 \cT^2 \left(\Xtz\right)^2
 + \left(1-\cT\right)^2 \left(1+2 \cT\right) \left(\Xtdm\right)^2
 + \left(1+\cT\right)^2 \left(1-2 \cT\right) \left(\Xtdp\right)^2
 \bigg]  \\
&  + \dfrac{3}{32} \, \bd  \bigg[   
   4 \left(1-\cT^2\right) \left(\Xtz\right)^2
 + \left(1-\cT\right)^3 \left(\Xtdm\right)^2
 + \left(1+\cT\right)^3 \left(\Xtdp\right)^2
 \bigg] \Bigg\} \ ,
\end{split}
\ee
\be
\llabel{211110t3}
\small
\begin{split}
\overline T_3 & = % \At \frac{\sf TMwc[1]}{\sT} = 
 - \At \sumk \Bigg\{ \dfrac{3}{32} \, \cT \, \az  \bigg[
   4 \left(1-3 \cT^2\right) \left(\Xtz\right)^2
 + 3 \left(1-\cT^2\right) \left(\left(\Xtdm\right)^2 + \left(\Xtdp\right)^2\right)
 \bigg]  \\
&  - \dfrac{3}{16} \, \au  \bigg[   
   4 \cT \left(1-2 \cT^2\right) \left(\Xtz\right)^2
 - \left(1-\cT\right)^2 \left(1+2 \cT\right) \left(\Xtdm\right)^2
 + \left(1+\cT\right)^2 \left(1-2 \cT\right) \left(\Xtdp\right)^2
 \bigg]  \\
&  + \dfrac{3}{32} \, \ad  \bigg[   
   4 \cT \left(1-\cT^2\right) \left(\Xtz\right)^2
 + \left(1-\cT\right)^3 \left(\Xtdm\right)^2
 - \left(1+\cT\right)^3 \left(\Xtdp\right)^2
 \bigg] \Bigg\} \ .
\end{split}
\ee
\bfx{\citet{Boue_etal_2016}\footnote{Note that the $b(\sigma)$ functions have a slightly different definition. In \citet{Boue_etal_2016} it is defined as the imaginary part of $\hat k_2 (\sigma)$, while in our case it is defined as the opposite of it (Eq.\,(\ref{210924d})).} 
obtained an equivalent expression for the double averaged tidal torque (Eq.(\ref{211110a})).
However, they only consider the contribution to the spin evolution (Eq.\,(\ref{210805b})), while here, we also apply the torque to the orbital evolution (Eq.\,(\ref{150626a})).
Moreover, to obtain the complete tidal evolution we additionally need to consider the evolution of the orbital energy (Eq.\,(\ref{211110t4})).}

\subsection{Orbital energy}

The evolution of the orbital energy is obtained %from the semi-major axis evolution (Eq.\,(\ref{211029f})), but it can also be obtained directly 
from the work of the tidal force $\vF$ (Eq.\,(\ref{170911d})) as
\be
\dot E_{\rm orb} = \big\langle \vF \cdot \dot \vr \big\rangle_{M,\vpi}  % = - \nabla U \cdot \dot \vr 
= -  \left\langle \frac{\partial U}{\partial t} \right\rangle_{M,\vpi} % = A_t \, {\sf TMwc[4]} 
\ . \llabel{211110b}
\ee
The first approach provides the Hansen coefficients in the format $X_k^{-4,m}$, which can be put into the format $X_k^{-3,m}$ using the relations provided in appendix~\ref{hcr}.
The second approach is easier to compute and already provides the Hansen coefficients in the format $X_k^{-3,m}$.
%As for the Laplace vector, after averaging we obtain the Hansen coefficients in the format $X_k^{-4,m}$, that can be put into the format $X_k^{-3,m}$ using the relations provided in appendix~\ref{hcr}.
Then,
\be
\llabel{211110t4}
\small
\begin{split}
\dot E_{\rm orb} & =
% n \At \frac{\sf TMwc[4]}{n} = 
 n \, \At  \sumk \Bigg\{ \dfrac{1}{64} \, k \,\bz  \bigg[
   4 \left(1-3 \cT^2\right)^2 \left(\Xtz\right)^2
 + 9 \left(1-\cT^2\right)^2 \left(\left(\Xtdm\right)^2 + \left(\Xtdp\right)^2\right)
 \bigg]  \\
&  + \dfrac{3}{16} \, k \, \bu \left(1-\cT^2\right) \bigg[   
   4 \cT^2  \left(\Xtz\right)^2
 +  \left(1-\cT\right)^2 \left(\Xtdm\right)^2
 + \left(1+\cT\right)^2 \left(\Xtdp\right)^2 
 \bigg]  \\
&  + \dfrac{3}{64} \, k \, \bd  \bigg[   
   4 \left(1-\cT^2\right)^2 \left(\Xtz\right)^2
 + \left(1-\cT\right)^4 \left(\Xtdm\right)^2 
 + \left(1+\cT\right)^4 \left(\Xtdp\right)^2
 \bigg] \Bigg\} \ .
\end{split}
\ee
% aqui In the limit of zero eccentricity, we have $X_k^{\ell,m} = \delta_{km}$ and this expression becomes equivalent to Eq.(28) in \citet{Lai_2012}.

\subsection{Orbital and spin evolution}
\llabel{oase2}

The set of equations (\ref{211110a}) and (\ref{211110t4}) allow us to track the evolution of the averaged system using the angular momentum vectors and the orbital energy.
As in Sect.~\ref{oase1}, we can relate these quantities with the orbital and spin parameters.
The semi-major axis is directly given from the orbital energy % (Eq.\,(\ref{211029e}))
\be
a = - \frac{\beta \mu}{2 E_{\rm orb}}
\ , \llabel{211110f}
\ee
and the eccentricity from the orbital angular momentum (Eq.\,(\ref{210804a}))
\be
e  % = \sqrt{1 - \frac{(\vG \cdot \vk)^2}{\beta^2 \mu a}}
= \sqrt{1 - \frac{\vG \cdot \vG}{\beta^2 \mu a}}
\ . \llabel{211110g}
\ee
The rotation rate, $\om$, is obtained from the rotational angular momentum (Eq.\,(\ref{211027c})), and the angle between the orbital and equatorial planes, $\theta$, from both angular momentum vectors (Eq.\,(\ref{211027z})).

As in Sect.~\ref{oase1}, we can also obtain the explicit evolution of all these quantities.
The semi-major axis evolution is given from expression (\ref{211110t4})
\be
\dot a = \frac{2 a^2}{\beta \mu} \, \dot E_{\rm orb} 
\ , \llabel{211110c}
\ee
while the eccentricity evolution can be computed from expressions (\ref{211110a}), (\ref{211110g}) and (\ref{211110c}) as
\be
\dot e =  \frac{1-e^2}{2 a e} \, \dot a  - \frac{\vG \cdot \vT}{\beta^2 \mu a e} 
=  \frac{\sq}{\beta n a^2 e} \left( \frac{\sq}{n} \, \dot E_{\rm orb} - \overline T_1 - \overline T_2 \, \cT \right)
\ , \llabel{211110d}
\ee
that is, using expressions (\ref{211110t1}), (\ref{211110t2}), and (\ref{211110t4}), 
\be
\llabel{211110e}
\small
\begin{split}
\dot e & = \Ae \, \frac{\sq}{e} \sumk \Bigg\{  \dfrac{1}{64} \, \bz  \bigg[ 
   4 \left(1-3\cT^2\right)^2 \left(\Xtz\right)^2 k \sq \\&
 + 9 \left(1-\cT^2\right)^2 \left(\Xtdm\right)^2 \left(2+ k \sq\right)  
 - 9 \left(1-\cT^2\right)^2 \left(\Xtdp\right)^2 \left(2- k \sq\right)  \bigg]  \\
 &  + \dfrac{3}{16} \, \bu  \bigg[   
    4 \left(1-\cT^2\right) \cT^2 \left(\Xtz\right)^2 k \sq  \\&
 + \left(1-\cT^2\right) \left(1-\cT\right)^2 \left(\Xtdm\right)^2 \left(2+k \sq\right)  
 - \left(1-\cT^2\right) \left(1+\cT\right)^2 \left(\Xtdp\right)^2 \left(2-k \sq\right)  \bigg]  \\
  & + \dfrac{3}{64} \, \bd  \bigg[   
 4 \left(1-\cT^2\right)^2 \left(\Xtz\right)^2 k \sq   \\&
 + \left(1-\cT\right)^4 \left(\Xtdm\right)^2 \left(2+k \sq\right)   
 - \left(1+\cT\right)^4 \left(\Xtdp\right)^2 \left(2-k \sq\right)  \bigg] \Bigg\} \ .
\end{split}
\ee
% aqui These expressions for the orbital evolution are equivalent to those obtained by \citet{Correia_Laskar_2010B} in the limit of small eccentricity.
For the rotation rate evolution, we have from expressions (\ref{210805b}) and (\ref{211027c}),
\be
\dot \om = - \frac{\vT \cdot \vs}{C} 
%= - \frac{T_s}{C} 
= - \frac{\overline T_1 \, \cT + \overline T_2}{C} 
\ , \llabel{211110h}
\ee
that is, using expressions (\ref{211110t1}) and (\ref{211110t2})
\be
\llabel{211110t5}
\small
\begin{split}
\dot \om & = % - \frac{1}{C} \At {\sf TMwc[3]} = 
- \frac{\At}{C} \sumk \Bigg\{ \dfrac{3}{16} \, \bu \left(1-\cT^2\right) \bigg[
   4 \cT^2  \left(\Xtz\right)^2
 + \left(1-\cT\right)^2  \left(\Xtdm\right)^2
 + \left(1+\cT\right)^2  \left(\Xtdp\right)^2
\bigg]  \\
&  + \dfrac{3}{32} \, \bd  \bigg[   
   4 \left(1-\cT^2\right)^2 \left(\Xtz\right)^2
 + \left(1-\cT\right)^4 \left(\Xtdm\right)^2
 + \left(1+\cT\right)^4 \left(\Xtdp\right)^2
 \bigg] \Bigg\} \ .
\end{split}
\ee
% aqui which is equivalent to the expression obtained by \citet{Dobrovolskis_1980, Correia_etal_2003, Cunha_etal_2015} in the limit of small eccentricity.
The obliquity (or inclination) evolution is given from expressions (\ref{211028z}) and (\ref{211110a}),
\be
\dot \theta  = \left( \frac{\overline T_1}{C \om} - \frac{\overline T_2}{\beta \sqrt{\mu a (1-e^2)}} \right)  \sin \theta
\ . \llabel{211110i}
\ee
Finally, for the angular velocity of the longitude of the node and for the precession speed of the spin axis, we get from expressions (\ref{211029a}), (\ref{211029b}) and (\ref{211110a}), respectively,
\be
\dot \Omega = \frac{\vT \cdot (\vk \times \vs)}{|\vG| \sin \theta} 
=  \frac{\overline T_3 \sin \theta}{\beta \sqrt{\mu a (1-e^2)}}
\ , \llabel{211110j}
\ee
\be
\dot \psi  = - \frac{\vT \cdot (\vk \times \vs)}{|\vL| \sin \theta} 
= - \frac{\overline T_3 \sin \theta}{C \om}
\ . \llabel{211110k}
\ee

% aqui: nao tenho a certeza que estes dois ultimos estao corretos, pois na expressao (102) e (103) aparecem $y$ e $z$ que interferem com a media. o valor correto deve obter-se diretamente de Tp medio (sem a subtracao de T4 e T5) 

\subsection{Energy dissipation}

The total energy released inside the body due to tides is given by expressions (\ref{211029d}) and (\ref{211029f}).
When we average over the argument of the pericentre, by combining expressions (\ref{211110t4}) and (\ref{211110t5}) we obtain for the total energy dissipated
\be
\llabel{211110l}
\small
\begin{split}
\dot E & =  \At  \sumk \Bigg\{ \dfrac{1}{64} \, (-k n) \,\bz  \bigg[
   4 \left(1-3 \cT^2\right)^2 \left(\Xtz\right)^2
 + 9 \left(1-\cT^2\right)^2 \left(\left(\Xtdm\right)^2 + \left(\Xtdp\right)^2\right)
 \bigg]  \\
&  + \dfrac{3}{16} \, (\om-k n) \, \bu \left(1-\cT^2\right) \bigg[   
   4 \cT^2  \left(\Xtz\right)^2
 +  \left(1-\cT\right)^2 \left(\Xtdm\right)^2
 + \left(1+\cT\right)^2 \left(\Xtdp\right)^2 
 \bigg]  \\
&  + \dfrac{3}{64} \, (2 \om -k n) \, \bd  \bigg[   
   4 \left(1-\cT^2\right)^2 \left(\Xtz\right)^2
 + \left(1-\cT\right)^4 \left(\Xtdm\right)^2 
 + \left(1+\cT\right)^4 \left(\Xtdp\right)^2
 \bigg] \Bigg\} \ .
\end{split}
\ee

\section{Planar case}

\llabel{planarcase}

The final outcome of tidal dissipation is the alignment of the spin axis with the normal to the orbit \citep{Hut_1980, Adams_Bloch_2015}.
Therefore, in order to simplify the equations of motion, many works assume that this alignment is always present, that is, the motion is planar ($\theta = 0$).
Indeed, in this case we have $\vk = \vs$ and thus (Eq.\,(\ref{211026a})):
\be
\cT %= \vk \cdot \vs 
= 1
 \ , \quad 
\es %= \ue \cdot \vs 
= 0
 \ , \quad \mathrm{and} \quad
\eks %= (\vk \times \ue) \cdot \vs 
= 0
\ . \llabel{211028g}
\ee

\subsection{Tidal torque}

Using the simplifications (\ref{211028g}) in expressions (\ref{211027t4}) and (\ref{211027t5}) yields $T_4 = T_5 = 0$.
In addition, since we also have $\vk \times \vs = 0$, we get for the average tidal torque (Eq.\,(\ref{211017d}))
%\be
%\vT = \frac{T_q}{\sin \theta} \, \vk + \left( T_s - \frac{T_q}{\sin \theta} \right) \vk + \frac{T_p}{\sin \theta} \, \vk \times \vk + 0 \, \ue + 0 \, \vk \times \ue = T_s \, \vk 
%\ee
\be
\left\langle \vT \right\rangle_M  = (T_1 + T_2) \, \vk = T_s \, \vk 
\ , \llabel{211028h}
\ee
with
\be
T_s %=  A_t {\sf TMab0[3]} =  
= \At \sumk  \dfrac{3}{2} \, \bd  \left(\Xtdp\right)^2 
\ . \llabel{211028j}
\ee
We note that, since $\es = \eks = 0$, the expression of the tidal torque is the same whether we perform a single average over the mean anomaly (Eq.\,(\ref{211017d})) or if we additionally average over the argument of the pericentre (Eq.\,(\ref{211110a})).

\subsection{Laplace vector}

Using the simplifications (\ref{211028g}) in expression (\ref{211026z2}) yields $\dot \ek = 0$.
The averaged Laplace vector then becomes (Eq.\,(\ref{211026b}))
\be
\de = \dot e \, \ue + \dot \vpi \, \vk \times \ve 
\ , \llabel{211029z}
\ee
with
\be
\small
\dot e  %= \Ae \, \frac{\sq}{e} \, {\sf dee0[1]}
=  \Ae \, \frac{\sq}{4 e} \sumk \Bigg\{ \bz  
   \left(\Xtz\right)^2 k \sq    - 3 \, \bd    
   \left(\Xtdp\right)^2 \left(2-k \sq\right) \Bigg\}
\ , \llabel{211029y}
\ee
and
\be
\small
\begin{split}
\dot \vpi %=  \Ae \, \frac{\sq}{e} \frac{1}{e} \,  \frac{\sf dee0[3]}{1-e^2}
 =   \frac{\Ae}{e^2 \sq} & \sumk \Bigg\{  \dfrac{3}{16} \, \az \bigg[
 2 e^2 (\Xtz)^2 + e^2 \Xtz \left(\Xtdm + \Xtdp\right) + 2 e \, \Xtz \left(\Xtum + \Xtup\right) \bigg]   \\
  & - \dfrac{1}{16} \, \ad \bigg[ 
 \left( 12 \left(2-k \left(1-e^2\right)^{3/2}\right)- 9 e^2\right) \left(\Xtdp\right)^2   + 3 e^2 \Xtdp \Xtdm \\&
 + \Big( 4 k \left(1-e^2\right)^{3/2} -6 e^2 \Big) \Xtz  \Xtdp  
 +6 e \,\Xtdp \left(\Xtum + \Xtup\right) \bigg] \Bigg\} \ .
\end{split}
\ee 
We note that in the planar case, $\dot e$ only depends on $b(\sigma)$, and $\dot \vpi$ only depends on $a(\sigma)$. 
%We thus easily see that tides modify the eccentricity on a secular timescale, but they only contribute for a precession of the pericentre.
Moreover, the expression of $\dot e$ does not change if we further average over the argument of the pericentre.

\subsection{Orbital and spin evolution}

For the semi-major axis, we have from expressions (\ref{211027e}), (\ref{211028j}) and (\ref{211029y}),
\be
\frac{\dot a}{a} %= \frac{2 \, \vG \cdot \vT}{\mu \beta^2 a (1-e^2)} + \frac{2 \, \ve \cdot \de}{(1-e^2)}
% = \frac{2 T_s}{\beta n a^2 \sqrt{1-e^2}} + \frac{2 e \dot e}{(1-e^2)} 
= \Ae \sumk  \frac{k}{2} \left[ \bz  
   \left(\Xtz\right)^2  + 3 \, \bd    
   \left(\Xtdp\right)^2 \right]
\ , \llabel{211103a}
\ee
while for the rotation rate, we get from expressions (\ref{211028a}) and (\ref{211028j}),
\be
\dot \om %= - \frac{\vT \cdot \vs}{C} 
%= - \frac{T_s}{C} 
= - \frac{\At}{C}  \sumk  \dfrac{3}{2} \, \bd  \left(\Xtdp\right)^2 
\ . \llabel{211103b}
\ee
Again, these expressions do not change if we further average over the argument of the pericentre.
As expected, the evolution of the obliquity (or inclination) is simply given by $\dot \theta =0$ (Eq.\,(\ref{211028z})), since $T_4 = T_5 = 0$ and $\sin \theta = 0$, that is, the motion remains planar.

\subsection{Energy dissipation}

The orbital and rotational energy variations can be obtained from expression (\ref{211029f}) together with expressions (\ref{211103a}) and (\ref{211103b}), respectively
\be
\dot E_{\rm orb} % = \frac{\beta \mu}{2 a^2} \dot a 
% = \beta n a^2 \frac{n}{2} \frac{\dot a}{a} 
% = \At \, {\sf TMab0[4]}
= \At \sumk  \frac{kn}{4} \left[ \bz  
   \left(\Xtz\right)^2  + 3 \, \bd    
   \left(\Xtdp\right)^2 \right]
\ , \llabel{211103c}
\ee
\be
\dot E_{\rm rot} =  - \At \sumk  \dfrac{3 \om}{2} \, \bd  \left(\Xtdp\right)^2 
\ . \llabel{211103d}
\ee
The total energy released inside the body due to tides is then (Eq.\,(\ref{211029d}))
\be
\dot E = \At \sumk \frac14  \left[ (-kn) \, \bz  
   \left(\Xtz\right)^2 + 3 \, (2 \om - kn) \, \bd    
   \left(\Xtdp\right)^2 \right]
\ . \llabel{211103e}
\ee
% aqui This expression is equivalent to Eq.(52) in \citet{Correia_etal_2014}.

\section{Linear model}

\llabel{linearapprox}

In Sect.~\ref{tidalmodels}, we present the tidal models most commonly used in the literature, and how they can be expressed in terms of the second Love number, $\hat k_2 (\sigma)$.
The linear or weak friction model is widely used because it is an approximation of any viscoelastic model for small relaxation times ($\sigma \tau \ll 1$) and provides simple expressions for the equations of motion that are valid for any eccentricity value.
Indeed, since $a(\sigma)$ is constant and $b(\sigma) \propto \sigma$ (Eq.\,(\ref{210929c})), the sum of the Hansen coefficients products can be evaluated as (see appendix~\ref{hcsum}):
\be
\sumk X_k^{-3,m} X_k^{-3,n} = X_0^{-6,m-n} 
\ , \label{211124a}
\ee
\be
\sumk k \, X_k^{-3,m} X_k^{-3,n} = \frac{\sq}{2}  (m+n) X_0^{-8,m-n}  
\ , \label{211124b}
\ee
\be
 \label{211124c}
 \begin{split}
\sumk k^2 \, X_k^{-3,m} X_k^{-3,n}  & = 9 \left( 2 X_0^{-9,m-n} - X_0^{-8,m-n} \right)  \\
& \quad + \left( \frac38 (m-n)^2 + m n - 9 \right) (1-e^2) X_0^{-10,m-n} \ ,
\end{split}
\ee
where the coefficients $X_0^{-\ell,m}$ can be obtained as exact functions of the eccentricity (Eq.\,(\ref{210910d}))
\be
\llabel{211128b}
\begin{split}
X_0^{-\ell,m} (e) & = \frac{1}{2 \pi} \int_{-\pi}^\pi \left( \frac{a}{r} \right)^\ell \ep^{\ii m \up} \, d M
= \frac{1}{\pi (1-e^2)^{\ell-3/2}} \int_0^\pi \left( 1+e \cos \up \right)^{\ell-2} \cos (m \up) \, d \up \\
& = \frac{1}{(1-e^2)^{\ell-3/2}} \sum_{k=0}^{(\ell-m-2)/2} \frac{(\ell-2)!}{k! (m+k)!(\ell-2-m-2k)!} \left( \frac{e}{2} \right)^{m+2k} \ .
\end{split}
\ee
We provide the explicit expression of all needed $X_0^{-\ell,m} (e)$ coefficients in Table~\ref{Hansen0}.
The equations of motion that we obtain here are in perfect agreement we those derived in \citet{Correia_2009} and \citet{Correia_etal_2011} using a different approach, and so for clearness reasons we keep the exact same notations for the eccentricity functions.

\begin{table*}
\begin{center}
\begin{tabular}{|ccc|ccc| } \hline 
$\ell$ & $m$ & $\bar X_0^{-\ell,m} (e) $ & $\ell$ & $m$ & $\bar X_0^{-\ell,m} (e) $ \\ \hline
$6$ & $0$ & $1 + 3 e^2 + \frac38 e^4$  &
$8$ & $0$ & $1 + \frac{15}{2} e^2 + \frac{45}{8} e^4 + \frac{5}{16} e^6$  \\
$6$ & $1$ & $2 e \, (1 + \frac34 e^2)$ &
$8$ & $1$ & $3 e \, (1 + \frac52 e^2 + \frac58 e^4)$  \\
$6$ & $2$ & $\frac32 e^2 \, (1 + \frac16 e^2)$  &
$8$ & $2$ & $\frac{15}{4} e^2 \, (1 + e^2 + \frac{1}{16} e^4)$  \\
$6$ & $3$ & $\frac12 e^3$  &
$8$ & $3$ & $\frac52 e^3 \, (1 + \frac38 e^2)$  \\
$6$ & $4$ & $\frac{1}{16} e^4$  &
$8$ & $4$ & $\frac{15}{16} e^4 \, ( 1 + \frac{1}{10} e^2)$  \\ \hline
$9$ & $0$ & $1 + \frac{21}{2} e^2 + \frac{105}{8} e^4 + \frac{35}{16} e^6$  &
$10$ & $0$ & $1 + 14 e^2 + \frac{105}{4} e^4 + \frac{35}{4} e^6 + \frac{35}{128} e^8$  \\
$9$ &$2$ &  $\frac{21}{4} e^2 \, (1 + \frac53 e^2 + \frac{5}{16} e^4) $  &
$10$ &$2$ &  $7 e^2 \, (1 + \frac52 e^2 + \frac{15}{16} e^4 + \frac{1}{32} e^6 )$  \\
$9$ & $4$ & $\frac{35}{16} e^4 \, (1 + \frac{3}{10} e^2) $ &
$10$ & $4$ & $\frac{35}{8} e^4 \, (1 + \frac35 e^2 + \frac{1}{40} e^4)$ \\ \hline
\end{tabular} 
\end{center}
\caption{Complete expression of the Hansen coefficients $X_0^{-\ell,m} = \bar X_0^{-\ell,m} / (1-e^2)^{\ell-3/2}$ (Eq.\,(\ref{211128b})), where $\bar X_0^{-\ell,m}$ is the polynomial part \citep[see also][Table~A.1]{Laskar_Boue_2010}). \llabel{Hansen0} }  
\end{table*}

\subsection{Average over the mean anomaly}

\subsubsection{Tidal torque}

The averaged tidal torque is given by expression (\ref{211017d}).
In the linear approximation, the coefficients $T_3 = T_5 = 0$, and thus, we obtain 
\be
\llabel{211130a}
\begin{split}
\big\langle \vT \big\rangle_M = \Kt \bigg[ & \left( \sq f_4(e) \frac{\om}{2n} \cos \theta  - f_2(e) \right) \vk 
%\\ & 
+ \left( f_1(e) - \frac12 \sq f_4(e) \right) \frac{\om}{n} \, \vs 
\\ & 
+ \left( \sq f_4(e)  - f_1(e) \right) \frac{\om}{n} \, (\ue \cdot \vs) \, \ue \, \bigg] \ , 
\end{split}
\ee
with
\be
f_1(e) = X_0^{-6,0} (e) = \frac{1 + 3 e^2 + \frac38 e^4}{(1-e^2)^{9/2}}
\ , \llabel{211130e}
\ee
\be
f_2(e) = \sq \, X_0^{-8,0} (e) = \frac{1 + \frac{15}{2} e^2 + \frac{45}{8} e^4 + \frac{5}{16} e^6}{(1-e^2)^6}
\ , \llabel{211130f}
\ee
\be
f_4 (e) = \frac{X_0^{-6,0} (e) - X_0^{-6,2} (e)}{\sq} = \frac{1 + \frac32 e^2 + \frac18 e^4}{(1-e^2)^5}
\ , \llabel{211130g}
\ee
and
\be
\Kt = 3 \, \At \,  \kf \, n \Delta t  = \kf \frac{3 \Gc \Ms^2 R^5}{a^6} \, n \Delta t 
\ . \llabel{211130h}
\ee

\subsubsection{Laplace vector}

The averaged Laplace vector is given by expression (\ref{211026b}).
In the linear approximation, we obtain 
\be
\llabel{211201a}
\begin{split}
\big\langle \de \big\rangle_M = \Ke \, \bigg[ & 
\left( 11 \,  f_4(e) \frac{\om}{2 n} \cos \theta - 9 \, f_5(e) \right) \ve 
%\\ & 
- f_4(e) \, \frac{\om}{2n} \, (\ve \cdot \vs) \, \vk  \, \bigg] 
%\\ & 
+  \kf \, \Ae \frac{15}{2} \, f_4(e) \, \vk \times \ve \ , 
\end{split}
\ee
with
\be
f_5 (e) = X_0^{-8,0} (e) - X_0^{-8,2} (e) = \frac{1 + \frac{15}{4} e^2 + \frac{15}{8} e^4 + \frac{5}{64} e^6}{(1-e^2)^{13/2}}
\ , \llabel{211201b}
\ee
and
\be
\Ke = 3 \, \Ae \, \kf \, n \Delta t  = \frac{\Kt}{\beta n a^2} 
= \kf \frac{3 \Gc \Ms^2 R^5}{\beta a^8} \, \Delta t  
\ . \llabel{211201c}
\ee

\subsubsection{Orbital evolution}

The evolution of the eccentricity and the argument of the pericentre in the linear approximation are already given by the first and third terms of the Laplace vector (Eq.\,(\ref{211201a})), respectively.
For the semi-major axis, we have from expression (\ref{211027e})
\be
\frac{\dot a}{a} = 2 \, \Ke \bigg[ f_2 (e) \frac{\om}{n} \cos \theta - f_3 (e) \bigg] 
 \ , \llabel{211201d}
\ee
with
\be
f_3 (e) =  6 X_0^{-9,0} (e) - 3 X_0^{-8,0} (e) - 2 (1-e^2) X_0^{-10,0} (e) = \frac{1 + \frac{31}{2} e^2 + \frac{255}{8} e^4 + \frac{185}{16} e^6 + \frac{25}{64} e^8}{(1-e^2)^{15/2}}
\ . \llabel{211201e}
\ee

\subsection{Average over the argument of the pericentre}

The eccentricity evolution (Eq.\,(\ref{211201a}), first term) and the semi-major axis evolution (Eq.\,(\ref{211201d})) do not depend on the pericentre.
In addition, the evolution of the pericentre (Eq.\,(\ref{211201a}), last term) does not depend on the dissipation ($\Delta t$), and therefore evolves on a shorter timescale than the orbit.
As a result, we can further average the equations of motion over the argument of the pericentre in order to get simpler expressions for the torque and spin evolution.

\subsubsection{Tidal torque}

The averaged tidal torque is now given by expression (\ref{211110a}).
In the linear approximation, it becomes
\be
\big\langle \vT \big\rangle_{M,\vpi} = \Kt \bigg[ f_1(e) \, \frac{\om}{2n} \left( \vs + \cos \theta \, \vk \right) - f_2(e) \, \vk \bigg] 
\ . \llabel{211201f}
\ee

\subsubsection{Spin evolution}

For the rotation rate, we get from expression (\ref{211110h}) 
\be
\dot \om = - \frac{\Kt}{C} \bigg[ f_1(e) \, \frac{\om}{2n} \left( 1 + \cos^2 \theta \right) - f_2(e) \cos \theta \bigg]  
\ , \llabel{211201f}
\ee
and for the obliquity (or inclination) evolution we have from expression (\ref{211110i}),
\be
\dot \theta = \frac{\Kt}{C \om} \bigg[f_1(e) \frac{\om}{2n} \cos \theta - f_2(e) \bigg] \sin \theta  
- \frac{\Ke}{\sqrt{1-e^2}} f_1(e) \frac{\om}{2n} \sin \theta 
\ . \llabel{211206a}
\ee

\subsection{Energy dissipation}

The orbital and rotational energy variations can be obtained from expression (\ref{211029f}) together with expressions (\ref{211201d}) and (\ref{211201f}), respectively.
%aqui
%\be
%\dot E_{\rm orb} = n \, \Kt \bigg[ f_2 (e) \frac{\om}{n} \cos \theta - f_3 (e) \bigg] 
%\ , \llabel{211206b}
%\ee
%\be
%\dot E_{\rm rot} = - \om \, \Kt \bigg[ f_1(e) \, \frac{\om}{2n} \left( 1 + \cos^2 \theta \right) - f_2(e) \cos \theta \bigg]  
%\ . \llabel{211206c}
%\ee
The total energy released inside the body due to tides is then (Eq.\,(\ref{211029d}))
\be
\dot E = n \, \Kt \bigg[ \frac12 \, f_1(e) \left( \frac{\om}{n} \right)^2 \left( 1 + \cos^2 \theta \right) - 2 \, f_2(e) \, \frac{\om}{n} \cos \theta + f_3(e)  \bigg] \ . \llabel{211206d}
\ee

\section{Conclusion}

\llabel{sectconc}

In this paper, we revisit the tidal evolution of a body disturbed by a point mass companion.
We derive the equations of motion in a vectorial formalism, where the basis depend only on the unit vectors of the spin and orbital angular momenta and on the Laplace unit vector.
These vectors are related to the spin and orbital quantities, thus easy to obtain and independent of the chosen frame.
We provide the expressions of the equations of motion for a single average over the mean anomaly, and also for an additional average over the argument of the pericentre.
We show that in both cases, the equations depend only on series of Hansen coefficients, that are widely used in celestial mechanics.
Our method is valid for any rheological model, which appears in the equations of motion through the second Love number. % , $\hat k_2(\sigma)$.
% aqui In the case of a linear model, we obtain an analytical result for the series of the Hansen coefficients and thus write exact expressions for the equations.

%In our model we assumed that $I_{ij} \ll C$ ($i,j=1,2,3$) and thus the rotational angular momentum is proportional to the angular velocity (Eq.\,(\ref{151019b})).
%The main implication is that $\vs$ is simply the unit vector of the rotational angular momentum.
%However, $\vs$ is actually the unit vector of the angular velocity, and so in a more accurate situation the spin evolution must be tracked by the evolution of the angular velocity.
%We have from expression (\ref{210804b}) 
%\be
%\dot \vw = - \TI' \cdot ( \vT + \dot \TI \cdot \vw )  \ , \quad \mathrm{with} \quad \TI' = ( C \, \mathbb{I} + \TI )^{-1}
%\ , \llabel{211208a}
%\ee
%where $\TI$ is given by expression (\ref{210910b}).

In our model, we use the quadrupolar approximation to obtain the tidal potential (Eq.\,(\ref{121026b})), that is, we neglect terms in $(R/r)^3$.
This approximation is usually suitable to study the long-term evolution of a large variety of systems, such as planet-satellite, planet-star or stellar binaries.
However, for extremely close-in bodies with very asymmetric shapes, such as Phobos (the main satellite of Mars) or binary asteroids, a high precision model of its tidal dynamics may require to include octupole or higher order terms in the tidal potential, as well as the knowledge of higher degree Love numbers \citep[e.g.,][]{Bills_etal_2005, Taylor_Margot_2010}.

In a more general two body problem, both bodies are expected to undergo tidal evolution.
As long as we keep the quadrupolar approximation, the cross terms of interaction can be neglected \citep[e.g.,][]{Boue_Laskar_2009}.
As a result, we only need to take into account a second contribution to the potential energy (Eq.\,(\ref{191014a})), where $\Ms$ is replaced by $\ms$, and $\TI$ is replaced by $\TI_0$ (which pertains to the body with mass $\Ms$).
Therefore, we get additional similar expressions for the tidal force (Eq.\,(\ref{170911d})) and for the torque (Eq.\,(\ref{151028e})), where $(\vp, \vq, \vs)$ are replaced by $(\vp_0, \vq_0, \vs_0)$ as they now also correspond to the body with mass $\Ms$.
The equations of motion for the spin of $\Ms$ are thus analogous to those for $\ms$ (Eq.\,(\ref{210805b})), while for the equations of motion for the orbit (Eqs.\,(\ref{150626a}) and (\ref{210805c})) we only need to add the contributions from the two bodies.

\bfx{The vectorial formalism presented in this paper is well suited to study the long-term evolution of celestial bodies. 
In addition to tidal effects, we usually need to consider the rotational deformation and general relativity corrections \citep[e.g.,][]{Correia_etal_2011}.
For multi-body systems, the secular interactions can be obtained either by developing the perturbing functions in terms of Legendre polynomials, suited for hierarchical systems \citep[e.g.,][]{Correia_etal_2016}, or in terms of Laplace coefficients, suited for non-resonant compact systems \citep[e.g.,][]{Boue_Fabrycky_2014a}.}

\begin{acknowledgements}
We thank G.~Bou\'e for discussions.
\bfx{We are grateful to two anonymous referees for their insightful comments.}
This work was supported by
CFisUC (UIDB/04564/2020 and UIDP/04564/2020),
GRAVITY (PTDC/FIS-AST/7002/2020),
PHOBOS (POCI-01-0145-FEDER-029932), and
ENGAGE SKA (POCI-01-0145-FEDER-022217),
%Enabling Green E-science for the SKA Research Infrastructure (ENGAGE SKA), 
funded by COMPETE 2020 and FCT, Portugal.
\end{acknowledgements}

\appendix

\section{Hansen coefficients relations}

\llabel{hcr}

From the definition of the Hansen coefficients (Eq.\,(\ref{210910c})), a number of recurrence relations can be obtained.
In this work we use the following ones \citep[e.g.,][]{Challe_Laclaverie_1969, Giacaglia_1976}:
\be
X_k^{\ell,-m} = X_{-k}^{\ell,m}
%X_k^{-3,-m} = X_{-k}^{-3,m}
\ , \llabel{210915b}
\ee
\be
% eq. (32) in Giacaglia_1976, comes from (a/r)^\ell = (a/r)^\{ell+1} (r/a) = (a/r)^\{ell+1} (1+e cosv)/(1-e^2)
(1-e^2) \, X_k^{\ell,m} = X_k^{\ell+1,m} + \frac{e}{2} \left[ X_k^{\ell+1,m-1} + X_k^{\ell+1,m+1} \right]
\ , \llabel{210915d}
\ee
\be
% eq. (29) in Giacaglia_1976, comes from the derivative
\llabel{210915c}
\begin{split}
\sq \, k \, X_k^{\ell,m}  & = m (1-e^2) X_k^{\ell-2,m}  + \frac{e \, \ell}{2 } \left[ X_k^{\ell-1,m-1} - X_k^{\ell-1,m+1} \right] \\
&= m \, X_k^{\ell-1,m} +  \frac{e}{2} \left[ (m+\ell) \, X_k^{\ell-1,m-1} + (m-\ell) \, X_k^{\ell-1,m+1} \right]
%e (m+3) \, X_k^{-4,m+1} + 2m \, X_k^{-4,m} + e (m-3) \, X_k^{-4,m-1} = 2 \sqrt{1-e^2} \, k \, X_k^{-3,m} 
\ . 
\end{split}
\ee

The first relation (Eq.\,(\ref{210915b})) allow us to obtain the coefficients $X_k^{-3,-1}$ and $X_k^{-3,-2}$ from the coefficients $X_k^{-3,1}$ and $X_k^{-3,2}$, respectively (Table~\ref{tabHansen}).
It also allow us to obtain any other coefficient with $m<0$ from $X_k^{\ell,m}$ with $m>0$.
The second relation (Eq.\,(\ref{210915d})) with $\ell = -4$ provides 
\be
\llabel{210924a}
\begin{split}
m=0 \quad \Rightarrow \quad (1-e^2) \, X_k^{-4,0} & = \frac{e}{2} \, X_k^{-3,-1} + X_k^{-3,0} + \frac{e}{2} \, X_k^{-3,1} \ , \\
 m=1 \quad \Rightarrow \quad (1-e^2) \, X_k^{-4,1} & = \frac{e}{2} \, X_k^{-3,0} + X_k^{-3,1} + \frac{e}{2} \, X_k^{-3,2} \ , \\
m=2 \quad \Rightarrow \quad (1-e^2) \, X_k^{-4,2} & = \frac{e}{2} \, X_k^{-3,1} + X_k^{-3,2} + \frac{e}{2} \, X_k^{-3,3} \ .
\end{split}
\ee
Finally, from the last relation (Eq.\,(\ref{210915c})) with $\ell = -3$ we get 
\be
\llabel{210924b}
\begin{split}
m=0 \quad \Rightarrow \quad \sqrt{1-e^2} \, k \, X_k^{-3,0}  & = \frac32 e \, ( X_k^{-4,1} - \, X_k^{-4,-1}) \ , \\
m=1 \quad \Rightarrow \quad \sqrt{1-e^2} \, k \, X_k^{-3,1} & = 2 e  \, X_k^{-4,2} + X_k^{-4,1} - e \, X_k^{-4,0} \ , \\
m=2 \quad \Rightarrow \quad \sqrt{1-e^2} \, k \, X_k^{-3,2} & = \frac52 e \, X_k^{-4,3} + 2 \, X_k^{-4,2} - \frac12 e \, X_k^{-4,1} \ .
\end{split}
\ee

Using these sets of relations, it is possible to express all Hansen coefficients appearing in this work solely as functions of $X_k^{-3,0}$, $X_k^{-3,1}$, and $X_k^{-3,2}$, using the following sequence:
\be
\llabel{210924c}
\begin{split}
X_k^{-4,3} &= \frac{1}{5e}\bigg[ e \, X_k^{-4,1} - 4 \, X_k^{-4,2} + 2 \sqrt{1-e^2} \, k \, X_k^{-3,2} \bigg] \ , \\
%X_k^{-4,-3} &= \left(e \, X_k^{-4,-1} - 4 \, X_k^{-4,-2} - 2 \sqrt{1-e^2} \, k \, X_k^{-3,-2} \right) / (5 e) \ , \\
X_k^{-3,3} &= \frac{1}{e} \bigg[ 2 (1-e^2) \, X_k^{-4,2} - 2 X_k^{-3,2} - e \, X_k^{-3,1} \bigg] \ , \\
%X_k^{-3,-3} &= \left( 2 (1-e^2) \, X_k^{-4,-2} - 2 X_k^{-3,-2} - e X_k^{-3,-1} \right) / e \ , \\
X_k^{-4,2} &= \frac{1}{2e} \bigg[ e \, X_k^{-4,0} - X_k^{-4,1} + \sqrt{1-e^2} \, k \, X_k^{-3,1} \bigg] \ , \\
%X_k^{-4,-2} &= \left( e \, X_k^{-4,0} - X_k^{-4,-1} - \sqrt{1-e^2} \, k \, X_k^{-3,-1} \right) / (2 e) \ , \\
X_k^{-4,1} &= \frac{1}{1-e^2}\bigg[ \frac{e}{2} \, X_k^{-3,2} +X_k^{-3,1}+\frac{e}{2} \, X_k^{-3,0} \bigg] \ , \\
%X_k^{-4,-1} &= \left( \frac{e}{2} \, X_k^{-3,0} + X_k^{-3,-1} + \frac{e}{2} \, X_k^{-3,-2} \right) / (1-e^2) \ , \\
X_k^{-4,0} &= \frac{1}{1-e^2} \bigg[ \frac{e}{2} \, X_k^{-3,-1} +X_k^{-3,0} +\frac{e}{2} \, X_k^{-3,1} \bigg] \ .
\end{split}
\ee

\section{Hansen coefficients combinations}

\llabel{hcsum}

%As in \citet{Correia_etal_2014}, we let 
We let
\be
\Fp = \left( \frac{r}{a} \right)^\ell \ep^{\ii m \up}
\ , \quad \mathrm{and} \quad
\Fm = \left( \frac{r}{a} \right)^\ell \ep^{- \ii n \up}
\ , \llabel{211124d}
\ee
with derivatives, respectively,
\be
\Fp' = - \ii \frac{d \Fp}{d M} = m \sq \left( \frac{r}{a} \right)^{\ell-2} \ep^{\ii m v} - e \, \ell \left( \frac{r}{a} \right)^{\ell-1} \frac{\ep^{\ii (m+1) \up} - \ep^{\ii (m-1) \up}}{2 \sq}
\ , \llabel{211124e}
\ee
\be
\Fm' = \ii \frac{d \Fm}{d M} = n \sq \left( \frac{r}{a} \right)^{\ell-2} \ep^{- \ii n \up} - e \, \ell \left( \frac{r}{a} \right)^{\ell-1} \frac{\ep^{-\ii (n+1) \up} - \ep^{-\ii (n-1) \up}}{2 \sq}
\ . \llabel{211124f}
\ee

Using the definition of the Hansen coefficients (Eq.\,(\ref{210910c})) we have
\be
\llabel{211124g}
\left\langle \Fp \Fm \right\rangle_M 
%= \left\langle \sumk X_k^{\ell,m} \, \ep^{\ii k M}  \sumk X_j^{\ell,-n} \, \ep^{\ii j M}  \right\rangle_M
= \sumk X_k^{\ell,m} X_k^{\ell,n} 
= \left\langle  \left( \frac{r}{a} \right)^{2\ell} \ep^{\ii (m-n) \up} \right\rangle_M
%= \left\langle  \sumk X_k^{2\ell,m-n} \, \ep^{\ii k M} \right\rangle_M
= X_0^{2\ell,m-n} \ , 
\ee
\be
 \llabel{211124h}
\begin{split}
\left\langle \Fp' \Fm \right\rangle_M 
% = \left\langle \sumk k \, X_k^{\ell,m} \, \ep^{\ii k M}  \sumk X_j^{\ell,-n} \, \ep^{\ii j M}  \right\rangle_M \\
& = \sumk k \, X_k^{\ell,m} X_k^{\ell,n} \\
&= \left\langle m \sq \left( \frac{r}{a} \right)^{2\ell-2} \ep^{\ii (m-n) \up} - \frac{e \, \ell}{2 \sq} \left( \frac{r}{a} \right)^{2\ell-1} \left( \ep^{\ii (m-n+1) \up} - \ep^{\ii (m-n-1) \up} \right)  \right\rangle_M \\
&= m \sq X_0^{2\ell-2,m-n} - \frac{e \, \ell}{2 \sq}  \left( X_0^{2\ell-1,m-n+1} - X_0^{2\ell-1,m-n-1} \right) \\
&= \frac{m+n}{2} \sq X_0^{2\ell-2,m-n} \ ,
\end{split}
\ee
\be
 \llabel{211124j}
\begin{split}
\left\langle \Fp' \Fm' \right\rangle_M
%& = \left\langle \sumk k \, X_k^{\ell,m} \, \ep^{\ii k M}  \sumk j \, X_j^{\ell,-n} \, \ep^{\ii j M}  \right\rangle_M \\
& = \sumk k^2 \, X_k^{\ell,m} X_k^{\ell,n} \\
&= \left\langle \frac{\ell^2 \, e^2}{4 (1-e^2)} \left( \frac{r}{a} \right)^{2\ell-2} \left( 2 \ep^{\ii (m-n) \up} - \ep^{\ii (m-n+2) \up} - \ep^{\ii (m-n-2) \up} \right)  \right\rangle_M \\
& \quad + \left\langle (m-n) \, \ell \, \frac{e}{2} \left( \frac{r}{a} \right)^{2\ell-3} \left( \ep^{\ii (m-n+1) \up} - \ep^{\ii (m-n-1) \up} \right) + m n \, (1-e^2) \left( \frac{r}{a} \right)^{2\ell-4} \ep^{\ii (m-n) \up}  \right\rangle_M \\
& = \frac{\ell^2 \, e^2}{4 (1-e^2)} \left( 2 X_0^{2l-2,m-n} - X_0^{2l-2,m-n+2} - X_0^{2l-2,m-n-2} \right) \\
& \quad + (m-n) \, \ell \, \frac{e}{2} \left( X_0^{2l-3,m-n+1} - X_0^{2l-3,m-n-1} \right) + m n \, (1-e^2) X_0^{2l-4,m-n} \\
& = \ell^2 \left( 2 X_0^{2l-3,m-n} - X_0^{2l-2,m-n} \right)  + \left( \frac{\ell (m-n)^2}{2\ell-2} + m n - \ell^2 \right) (1-e^2) X_0^{2l-4,m-n} \ ,
\end{split}
\ee
where to simplify expression (\ref{211124h}) we used equation (\ref{210915c}), and to simplify expression (\ref{211124j}) we used equations (\ref{210915d}) and (\ref{210915c}) together with \citep[eg.][]{Giacaglia_1976}:
\be
(1-e^2)^2 X_k^{\ell,m} = \left(1+\frac{e^2}{2}\right) X_k^{\ell+2,m} + e \left[ X_k^{\ell+2,m+1} + X_k^{\ell+2,m-1} \right] + \frac{e^2}{4} \left[ X_k^{\ell+2,m+2} + X_k^{\ell+2,m-2} \right] 
\ . \llabel{211128a}
\ee

\bibliographystyle{spbasic}      % basic style, author-year citations
\bibliography{\bibpath correia}
  % name your BibTeX data base

\end{document}